\documentclass[12pt,letter]{article}
\pdfoutput=1
\usepackage{graphicx, epsfig, color,cite}
\usepackage{amsmath}
\usepackage{amssymb}
\usepackage{float}
\usepackage{subfig}
\usepackage{hyperref}

\def\to{\rightarrow}

\def\bi{\begin{itemize}}
\def\ei{\end{itemize}}

\def\tchi{\tilde\chi}

\def\tf{\tilde f}

\def\tst{\tilde t}

\def\tg{\tilde g}

\def\tq{\tilde q}
\def\alt{\lesssim}
\def\agt{\gtrsim}
\def\be{\begin{equation}}  
\def\ee{\end{equation}}  
\def\bea{\begin{eqnarray}}  
\def\eea{\end{eqnarray}}

\begin{document}
\begin{titlepage}
\begin{flushright}
OU-HEP-260102
\end{flushright}

\vspace{0.5cm}
\begin{center}
      {\Large \bf pMSSM versus complete models and the excellent\\ prospects for top-squark discovery at HL-LHC}\\
\vspace{1.2cm} \renewcommand{\thefootnote}{\fnsymbol{footnote}}
{\large Howard Baer$^{1}$\footnote[1]{Email: baer@ou.edu },
  Vernon Barger$^2$\footnote[2]{Email: barger@pheno.wisc.edu},
  and Kairui Zhang$^1$\footnote[4]{Email: kzhang25@ou.edu}
}\\ 
\vspace{1.2cm} \renewcommand{\thefootnote}{\arabic{footnote}}
{\it 
$^1$Homer L. Dodge Department of Physics and Astronomy,
University of Oklahoma, Norman, OK 73019, USA \\[3pt]
}
{\it 
$^2$Department of Physics,
University of Wisconsin, Madison, WI 53706 USA \\[3pt]
}
\end{center}

\vspace{0.5cm}
\begin{abstract}
\noindent
LHC sparticle search limits are usually performed within the context of simplified models and subsequently interpreted within the 19 parameter
phenomenological MSSM (pMSSM) as to how many models avoid search limits 
for a particular sparticle mass, often including WIMP dark matter 
constraints. 
We provide a critical discussion of this procedure and how it can go
wrong due to the introduction of new prejudices. 
By ameliorating these conditions, one is pushed into the more plausible
four extra parameter non-universal Higgs model (NUHM4). 
Implementing a decoupling/quasi-degeneracy solution to the SUSY flavor and CP problems leads to first/second generation sfermions in the tens-of-TeV range. 
In this case, the natural solutions typically contain top-squarks in the
1-2 TeV range which are accessible to high-lumi LHC (HL-LHC) searches. 
This search channel, along with higgsino and wino pair production, may allow a nearly complete scan of natural/plausible parameter space by HL-LHC.

\end{abstract}
\end{titlepage}

\section{Introduction}
\label{sec:intro}

Weak scale supersymmetry (SUSY)\cite{Baer:2006rs,Baer:2020kwz} is highly motivated in that it provides an elegant solution to the Higgs mass instability problem which is inherent in the Standard Model (SM). 
A remnant $N=1$ spacetime SUSY emerges from string compactification 
on Calabi-Yau manifolds owing to the properties of special holonomy\cite{Candelas:1985en}. 
It is conjectured that CY compactifications are the only stable string compactifications\cite{Acharya:2019mcu}.
If so, then the remaining question is: at which mass scale is SUSY broken?
Whereas the magnitude of the weak scale is put in by hand in the SM, 
it arises instead as a derived quantity as a consequence of the 
values of the soft SUSY breaking terms in the MSSM. This leads one to expect sparticle masses of order the weak scale which are then amenable to searches at the CERN 
Large Hadron Collider (LHC).

LHC sparticle search analyses usually take place within the framework of
simplified models which assume particular production processes along with
single or no sparticle decay modes. To gain context, 
the search results are often interpreted within the framework of the
phenomenological MSSM (pMSSM) model\cite{MSSMWorkingGroup:1998fiq} which assumes 19 independent 
parameters defined at the weak scale. 
Use of the pMSSM is invoked in that it may be general enough to encompass a wide variety of possible theoretical scenarios. 
pMSSM search results are often presented in terms of the 
ratio of number of scanned models which survive experimental search limits
compared to the total number of pMSSM models that were scanned over.
Several recent reviews include final sections on pMSSM analyses\cite{Dickinson:2022hus,ATLAS:2024lda,Sekmen:2025bxv,Constantin:2025mex}, and
are able to conclude that the bulk of SUSY model parameter space is now excluded although viable portions still remain, and so WSS is 
mostly but not yet fully excluded.

An alternative approach to simplified models and pMSSM analyses is to appeal to more complete SUSY models that include certain theoretical and phenomenological motivations (which may be termed as theoretical 
{\it prejudice}\cite{Berger:2008cq}). 
These are usually based on some type of communication of SUSY breaking from the hidden sector to the visible sector, such as gravity mediation\cite{Chamseddine:1982jx,Barbieri:1982eh,Ohta:1982wn,Hall:1983iz} (SUGRA), gauge mediation\cite{Dine:1995ag,Giudice:1998bp} (GMSB), anomaly mediation\cite{Randall:1998uk,Giudice:1998xp} (AMSB) or gaugino mediation\cite{Schmaltz:2000gy} (inoMSB). There are also models of mixed mediation such as mirage mediation (MM) where gravity and anomaly-mediated
contributions to soft terms are comparable\cite{Choi:2005ge}.
In several of these scenarios-- GMSB, AMSB and inoMSB-- the small trilinear
$A$ terms (which can otherwise boost the light Higgs mass into the
$m_h\sim 125$ GeV range with TeV-scale top-squarks\cite{Carena:2002es,Baer:2011ab,Arbey:2011ab,Hall:2011aa,Baer:2024fgd}) require large, 
unnatural top-squark soft terms in order to explain the rather large 
measured value of $m_h$\cite{ATLAS:2012yve,CMS:2012qbp}. 
Thus, these models, at least in their minimal incarnations, seem rather implausible if not excluded\cite{Baer:2014ica}, in light of 
LHC Higgs mass measurements. 
In contrast, in gravity mediation, one
generically expects large, TeV-scale trilinear soft terms. 
We take this to indicate that gravity-mediation is selected out by
LHC data as the remaining most plausible SUSY model\cite{Baer:2024fgd}. For this reason, 
here we will focus on the original SUGRA models or MM models as the most plausible of the extant SUSY models.

In the present paper, we delineate a number of flaws in pMSSM analyses which can lead to errant conclusions concerning the viability of WSS and the most plausible routes to discovery. 
These include the following.
\bi
\item Generic non-universality of gaugino masses
\item Non-unified matter scalar masses in spite of the fact that the matter superfields fill out the 16-dimensional spinor representation of $SO(10)$.
Indeed, Nilles {\it et al.}\cite{Nilles:2014owa} include matter in spinors of $SO(10)$ as amongst their ``golden rules'' of SUSY phenomenology.
\footnote{Nilles' five golden rules of string phenomenology: 1. matter scalars in spinors of $SO(10)$, 2. incomplete gauge and Higgs multiplets, 3. repetition of families, 4. $N=1$ SUSY and 5. $R$-parity and other discrete symmetries.}
\item Lack of renormalization group evolution of gauge and Yukawa couplings and soft SUSY breaking terms. This aspect abandons the successes of gauge coupling unification\cite{Ellis:1990wk,Amaldi:1991cn,Langacker:1991an} and radiative electroweak symmetry breaking (REWSB)\cite{Ibanez:1982fr,Alvarez-Gaume:1983drc} via the large top-quark Yukawa coupling.
\item Artificial upper bounds on pMSSM scan limits, and log versus uniform sampling of pMSSM soft terms.
\item The notion of electroweak naturalness. 
Quantifying naturalness via conventional measures\cite{Baer:2013gva} is dubious or not possible for the pMSSM, but for the newer, more conservative measure 
$\Delta_{EW}$\cite{Baer:2012up}, naturalness depends only on the weak scale mass spectrum,
and so for a given weak scale sparticle/Higgs mass spectrum, 
the same value of $\Delta_{EW}$ is obtained whether the spectrum is 
generated from some high scale model or from the pMSSM.
\item The method invoked to plausibly solve the SUSY flavor and CP problems.
The pMSSM invokes a degeneracy solution to the SUSY flavor/CP problem
by setting first and second generation soft terms equal to each other. 
However, such degeneracy is in contradiction to fundamental expectations from gravity mediation\cite{Soni:1983rm,Kaplunovsky:1993rd,Brignole:1993dj}. 
An alternative\cite{Dine:1993np,Cohen:1996vb,Baer:2010ny} is the 
decoupling, quasi-degeneracy solution which arises from fundamental considerations of the string landscape\cite{Baer:2019zfl}.
\item In many pMSSM analyses, the authors elect to invoke constraints 
from WIMP dark matter (DM) (see {\it e.g.} \cite{vanBeekveld:2016hug}). 
Whereas this seemed to be a paradigm DM model many years ago\cite{Jungman:1995df}, nowadays it is understood that there are a variety of mechanisms that can augment or suppress the neutralino DM relic density\cite{Gelmini:2006pw}, that WIMPs may occur alongside other well-motivated DM candidates such as the axion\cite{Baer:2011hx}, or even plausible scenarios for all axion/ no WIMP dark matter\cite{Baer:2025oid}. 
Invoking DM constraints can surely lead one astray. 
\ei

In Sec. \ref{sec:pmssm}, we discuss each of these issues in light of the present LHC interpretations in terms of pMSSM scans. Next, we show in Sec. \ref{sec:scan} numerically how different pMSSM scans can lead to different conclusions regarding SUSY, and as to the importance of different SUSY search strategies. 
In Sec. \ref{sec:nuhm4}, we implement corrections to these issues, which then places us in the four-extra-parameter non-universal Higgs model NUHM4
with a mixed decoupling/quasi-degeneracy solution to the SUSY flavor/CP
problems. 
This model, unlike the CMSSM/mSUGRA model, should be the expected low energy effective field theory (LE-EFT) of gravity mediation within the MSSM.
Some of the lessons this model has for SUSY discovery physics 
can be displayed in the $m_0(3)$ vs. $m_{1/2}$ parameter plane. 
From this plane, it is easy to see why it would have been surprising 
had SUSY already been discovered at LHC in Run 2 data: the bulk of LHC-accessible parameter-space actually lies within the charge-or-color-breaking (CCB) excluded region. 
It also suggests the promise of 
top-squark pair production searches for LHC Run 3 and HL-LHC in that 
the natural regions of parameter space all have $m_{\tst_1}\alt 2$ TeV 
whilst the HL-LHC reach extends to nearly saturate this limit.
Combined with pair production of light higgsinos, stop pair production
offers excellent prospects for SUSY discovery at LHC in the coming years.

 \section{Some issues in pMSSM modeling}
\label{sec:pmssm}
 
\subsection{Gaugino mass unification}
\label{ssec:GMU}

In SUSY models derived from string compactification,
the gauge couplings arise from the vacuum value of the dilaton modulus.
The canonically normalized gaugino masses are given by the term\cite{Brignole:1993dj}
\be
M_a ={1\over 2}(Re\ f_a)^{-1} F^m\partial_m f_a
\ee
where $f_a(h_m)$ is the holomorphic gauge kinetic function and $h_m$
denotes a (collection of) hidden sector superfield(s) responsible for
SUSY breaking.
$F_m$ is the $F$-term of the hidden sector
field $h_m$ which gains a SUSY-breaking vev $F^m\sim m_{hidden}^2$ where
$m_{hidden}\sim 10^{11}$ GeV to generate weak scale scale soft terms and
$\partial_m\equiv \partial /\partial h_m$.
To generate gaugino masses of order $m_{weak}\sim m_{3/2}$, then $f_a$
must be a non-trivial function of hidden sector fields and in SUGRA
from string models has a leading contribution $f_a =k_a S$, where
$S$ is the dilaton superfield giving rise also to the gauge coupling
and $k_a$ is the Kac-Moody level.
In this form, {\it universal} gaugino masses of order $m_{3/2}$ are
obtained at the unification scale even without any grand unification.

Arbitrary gaugino masses can arise in intersecting $D$-brane string models\cite{Blumenhagen:2005mu}, but these also usually contradict gauge coupling unification unless some recondite physics is invoked.

Non-universal gaugino masses can be obtained in SUGRA models in cases
where one includes loop contributions which
assume the anomaly-mediated SUSY breaking form\cite{Randall:1998uk,Giudice:1998xp}:
\be
M_a(1\text{-loop})\simeq -\frac{b_a g_a^2}{16\pi^2}m_{3/2} 
\ee
where the $b_a$ are coefficients of the RGEs of the $g_a$ gauge couplings with $b_a=(-11,-1,3)$.
For $M_a\sim m_{3/2}$ at tree level, then the AMSB corrections are of order $0.5$\% of the tree level approximation and so provide only a 
tiny correction to gaugino mass universality. 
But under some form of sequestering, then the tree-level $M_a$ may be suppressed so that the AMSB loop contributions
are comparable (mirage mediation) or even dominant (AMSB).

For example, gaugino masses in SUGRA arise via
\be
\int d^2\theta \frac{Y W^\alpha W_\alpha}{m_P}\to \frac{F_Y}{m_P}\lambda\lambda
\label{eq:mino}
\ee
leading to weak scale gaugino masses when the hidden sector field $Y$
attains a SUSY breaking vev $F_Y$. It is important to notice here
that the SUSY breaking field $Y$ must be a gauge singlet and further that
generation of gaugino masses leads to a broken $R$-symmetry in
the Lagrangian since the gauginos carry $R$-charge $+1$.
Trilinear terms arise as
\be
\int d^2\theta \frac{Y H_uQ U^c}{m_P}\to \frac{F_Y}{m_P} H_u\tilde{Q}\tilde{u}_R
\label{eq:Aterms}
\ee
{\it etc.} so that trilinear soft terms $a_{ijk}\sim m_{weak}$ also
develop.
As for gaugino masses, the field $Y$ must be a hidden sector singlet.

The rather large value of $m_h\simeq 125$ GeV can naturally arise provided
there are $A$-terms of order the weak scale\cite{Carena:2002es,Baer:2011ab,Arbey:2011ab,Baer:2024fgd}.
In this case, natural values of top squark masses  $m_{\tst_1}\sim 1-3$ TeV
and $m_{\tst_2}\sim 3-8$ TeV are allowed.
In contrast, for suppressed $A$-terms, then top-squark masses in the
$m_{\tst_i}\sim 10-100$ TeV are needed.
These provide TeV-scale corrections $\Sigma_u^u(\tst_{1,2})$ to the weak scale,
resulting in the need for an implausible finetuning.
Running the argument in reverse, we can say that the large Higgs mass implies
large tree-level SUGRA $A$-terms which then imply unsequestered,
universal gaugino masses.

\begin{figure}[htb!]
\centering
    {\includegraphics[height=0.5\textheight]{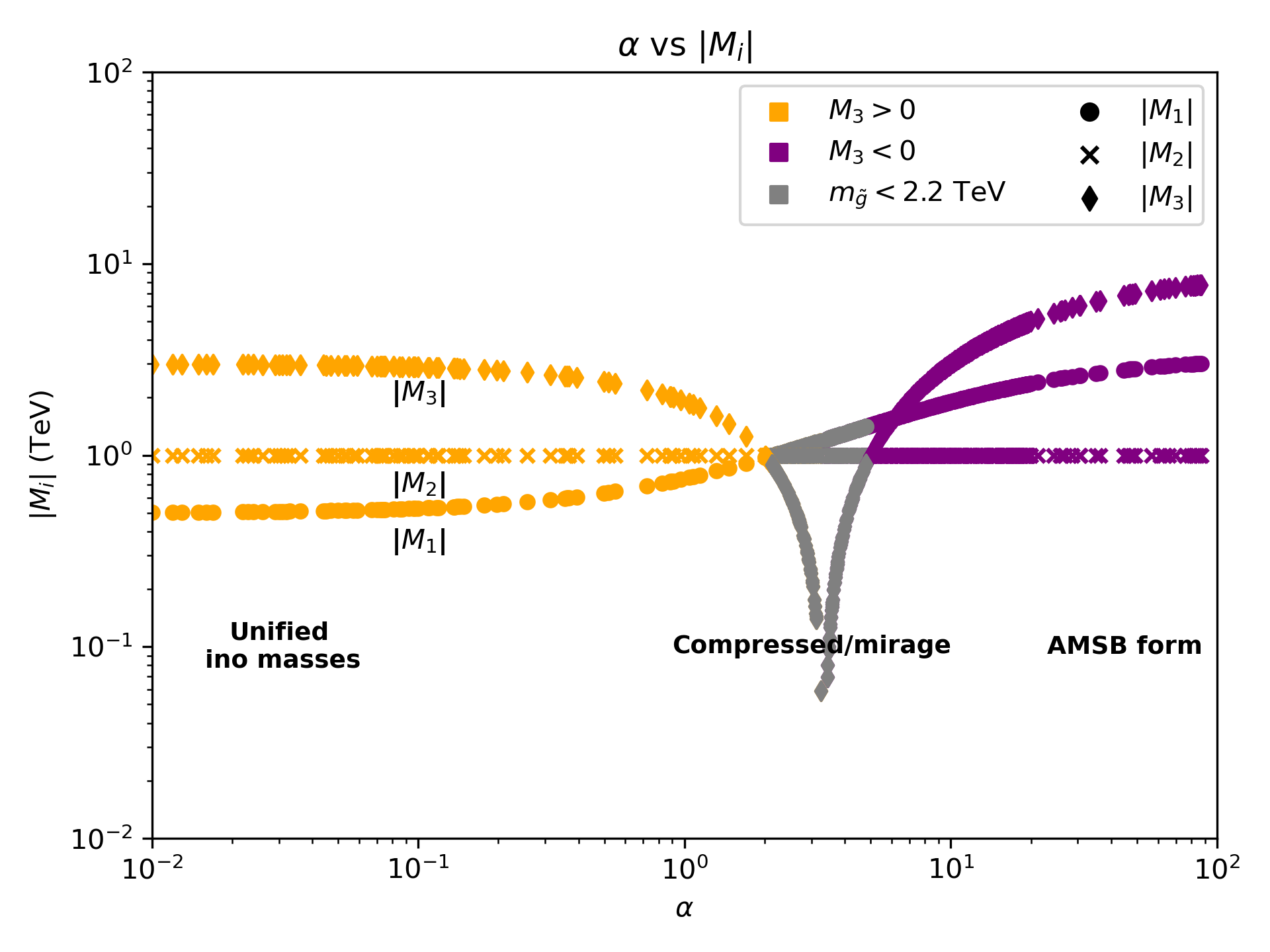}}
        \caption{Plot of weak scale gaugino masses vs. 
        mirage mediation mixing parameter  $\alpha$.
      \label{fig:Mi}}
\end{figure}

The form of gaugino masses expected at the weak scale is plotted
in Fig. \ref{fig:Mi} vs. $\alpha$, where $\alpha$ is an arbitrary parameter
that interpolates between universal gaugino masses for $\alpha\sim 0$
(which give rise to the ratio $M_1:M_2:M_3\sim 1:2:6$ at the weak scale),
and the expected form from AMSB (the limit for large $\alpha $):
\be
M_1:M_2:M_3\sim (1+0.66\alpha :\ 2+0.2\alpha :\ 6-1.8\alpha )
\ee
The ratios are plotted assuming a fixed value of wino mass $M_2 =1$ TeV.
In this form, pMSSM gaugino mass configurations such as $M_{1,2}\gg\mu$
which gives rise to sub-GeV higgsino mass gaps never arise.

\subsection{Matter unification in spinors of $SO(10)$}
\label{ssec:spinor}

It is a rather generic feature of gravity-mediated SUSY breaking models that
scalar masses are expected to be non-universal\cite{Soni:1983rm,Kaplunovsky:1993rd,Brignole:1993dj}.
This especially holds true in realistic string models with compactification  on a Calabi-Yau manifold, 
where of order hundreds of moduli fields may be
present\cite{Baer:2020vad}.
Such scalar mass non-universality typically leads to large contributions to
flavor-changing neutral current processes, and indeed has provided a strong
motivation for development of flavor-conserving models with generational
universality such as gauge mediation (GMSB)\cite{Dine:1995ag} and AMSB\cite{Randall:1998uk,Giudice:1998xp}.
The assumed scalar mass universality present in models like CMSSM\cite{Kane:1993td}
made these {\it unlikely} frameworks to arise from gravity-mediation.

However, one form of scalar mass universality is highly motivated: that is
universality of all scalar masses within each generation. This is because
the 16 matter superfields of each generation fit snugly into the 16-dimensional
spinor representation of $SO(10)$ SUSY GUTs. 
This is especially motivated
by the emergence of incontrovertible data on neutrino masses which requires
the addition of a superfield $N^c_i$ containing a (conjugate)
right-hand neutrino for each generation $i=1-3$.
The $N^c_i$ can gain high-scale Majorana masses from the superpotential
$W\ni{1\over 2}M_{ij}N_i^cN_j^c$ leading to see-saw Majorana neutrino masses\cite{Gell-Mann:1979vob,Yanagida:1979as}. 
It is also highly motivated by the (almost magical) cancellation of
axial anomalies in the SM and the MSSM.
In contrast, under $SO(10)$ unification the magic disappears and
anomaly cancellation is simply a by-product of $SO(10)$ unification\cite{Wilczek:1981iz}. 
Now, grand unification in four dimensions has its own set of
troubles, part of which is that the large Higgs representations needed for
spontaneous GUT symmetry breakdown tend to not emerge from string
compactifications.
Here we have in mind instead the notion of local GUT unification\cite{Buchmuller:2005sh,Nilles:2009yd}
where some multiplets such as the 16 occur due to the corresponding field location on  the compactified space. Other fields such as Higgs and gauge may occur in split multiplets since they lie in different locales in the compactified space.
Indeed, matter multiplets in spinors of $SO(10)$ comprise one of Nilles'
five golden rules of string phenomenology\cite{Nilles:2004ej}.
Thus, we expect in gravity mediation that the high scale sfermion masses obey
\be
m_{Q_i}=m_{U^c_i}=m_{D^c_i}=m_{L_i}=m_{E^c_i}\equiv m_0(i)
  \ee
  with
  \be
  m_0(1)\ne m_0(2)\ne m_0(3)\ne m_{H_u}\ne m_{H_d} .
  \ee

  In this case, one might expect the SUSY flavor problem to persist
  due to the generational matter non-universality.
  However, under the string landscape, the soft terms are expected to be
  statistically preferred to be at
  large values but subject to the (anthropic) requirement that they not
  contribute more than a factor of a few to the measured value of the weak
  scale (ABDS\cite{Agrawal:1997gf} window). 
  The upper bound for matter generations 1 and 2 is
  generation-independent and arises from 2-loop RGE contributions\cite{Martin:1993zk} that
  tend to drive the top-squark soft terms tachyonic\cite{Arkani-Hamed:1997opn}.
  This occurs around $m_0(i)\sim 20-50$ TeV\cite{Baer:2017uvn},
  leading to a landscape-generating decoupling/quasi-degeneracy solution
  to the SUSY flavor (and CP) problems\cite{Baer:2019zfl}
  (for previous earlier related work, see {\it e.g.} \cite{Dine:1990jd,Dine:1993np,Cohen:1996vb}).
  Thus, we expect from gravity-mediated SUSY which arises from
  the string landscape to have
  \be
  m_0(3)\ll m_0(1)\sim m_0(2)\sim 20-50\ {\rm TeV}.
  \ee
  We note here that the $SO(10)$ matter symmetry also ameliorates
  the most stringent FCNC bounds arising from the kaon sector\cite{Moroi:2013sfa}. Given this situation, the expectation is that first/second generation
  matter scalars should largely decouple from LHC (and other) SUSY search processes.

  A further advantage to arranging sfermions in spinors of $SO(10)$ is the
  possibility of increased naturalness\cite{Baer:2013jla}.
  Neglecting Yukawa couplings for first/second generation sfermions, their contribution to the naturalness measure
  $\Delta_{EW}$ from   Coleman-Weinberg loop corrections\cite{Coleman:1973jx} is
  \be
  \Sigma_{u,d}^{u,d}(\tf_{L,R})=\mp\frac{c_{col}}{16\pi^2}F(m_{\tf_{L,R}}^2)(-4g_Z^2(T_3-Q_{em}x_W))
  \ee
  where $T_3$ is weak isospin, $Q_{em}$ is electric charge, $c_{col}=1$ (3) for
  sleptons (squarks), $x_W=\sin^2\theta_W$ and
  $F(m^2)=m^2\left(\log\frac{m^2}{Q^2}-1\right)$. 
  The optimized scale choice $Q^2=m_{\tst_1}m_{\tst_2}$. 
  By summing over a complete generation of squarks and sleptons,
  the values of terms involving $T_3$ and $Q_{em}$ largely cancel out so that these terms
  give only a tiny contribution to $\Delta_{EW}$.
  
\subsection{The case for renormalization group evolution (RGE)}
\label{ssec:RGEs}

A further peculiarity of using the pMSSM model is the abandonment of
renormalization group evolution of soft terms.
This means in particular the discarding of the major success of MSSM gauge
coupling unification. 
One may stipulate that gauge couplings should still
(nearly) unify if the pMSSM is embedded into a high scale theory, but even so,
the gauge coupling evolution enters directly into the evolution of all
other soft terms. 

Another major success which is abandoned in the pMSSM
is the radiative-driving of electroweak gauge symmetry breaking (REWSB)
by the large
top-quark Yukawa coupling\cite{Ibanez:1982fr,Alvarez-Gaume:1983drc}.
The requirement of successful EWSB feeds into the evolution of all the 
other soft terms. 
It also feeds into certain values of soft terms which lead to charged or colored LSPs or charge or color breaking (CCB) vacua\cite{Casas:1995pd}.
This is of special importance in the case of 10-50 TeV first/second generation
matter scalars in that the favored low values of $m_0$ and $m_{1/2}$ in models
such as the CMSSM are all excluded by the large sfermion mass values 
which-- via 2-loop RG effects-- drive top-squark soft terms to tachyonic values\cite{Arkani-Hamed:1997opn}.
This effect can be interpreted as a reason why SUSY with several hundred GeV
soft terms has not been discovered at LHC\cite{Baer:2024hpl}.
One may keep the large 19-dimensional parameter dependence of the SUSY model but without abandoning the RGEs\cite{Baer:2013bba}.

\subsection{The naturalness issue}
\label{ssec:nat}

In the MSSM, the magnitude of the weak scale is actually derived 
in terms of the soft SUSY breaking terms such that
\be
m_Z^2/2=\frac{m_{H_d}^2+\Sigma_d^d-(m_{H_u}^2+\Sigma_u^u)\tan^2\beta}{\tan^2\beta -1}
  -\mu^2\simeq -m_{H_u}^2-\mu^2-\Sigma_u^u(\tst_{1,2})
  \label{eq:mzs}
\ee
where $m_{H_u}^2$ and $m_{H_d}^2$ are soft SUSY breaking Higgs masses,
$\mu$ is the (SUSY-conserving) Higgs/higgsino superpotential mass term\footnote{Twenty solutions to the SUSY $\mu$ problem are reviewed in Ref. \cite{Bae:2019dgg}.},
$\tan\beta \equiv v_u/v_d$ is the ratio of Higgs field vacuum expectation values
and the $\Sigma_u^u$ and $\Sigma_d^d$ terms contain an assortment of radiative
corrections, the most important of which are usually the
$\Sigma_u^u(\tst_{1,2})$\cite{Baer:2012cf}.
In accord with practical naturalness--
wherein all independent contributions to an observable are comparable to or less than that observable-- then, roughly, soft SUSY breaking terms should be of order the TeV scale.

A numerical measure of the naturalness of the weak scale in the MSSM is
given by 
\be
\Delta_{EW}\equiv max_i| i^{th}\ term\ on\ RHS\ of\ Eq.~\ref{eq:mzs}|/(m_Z^2/2) .
\ee
A value of $\Delta_{EW}\alt 30$ requires (the square root of the absolute value of) each independent contribution
to $m_Z^2/2$ to be within a factor of four of $m_Z$ itself.
Quantification of naturalness can be an abstract concept and is sometimes
criticized as being subjective.
A perhaps more concrete way of thinking about it is that,
{\it without finetuning},
SUSY models predict the magnitude of the weak scale to be
\be
m_{weak}\equiv m_{W,Z,h}\simeq 2 \sqrt{max_i| i^{th}\ term\ on\ RHS\ of\ Eq.~\ref{eq:mzs}|} .
\ee
In this context, SUSY models that give a bad prediction of the
measured value of the weak scale are those that are finetuned,
whilst those that give a good prediction are natural.
A rough boundary between the two occurs around $\Delta_{EW}\sim 30$.

Aside from being more conservative than other measures, an advantage of the $\Delta_{EW}$ measure is that it is uniquely determined by the weak scale sparticle mass spectrum. Thus, whether the spectrum is derived from the pMSSM or some high scale model, it will have the same naturalness value of $\Delta_{EW}$. This is not true of other measures\cite{Baer:2013gva}.

\subsection{Solution to the flavor/CP problems}
\label{ssec:flavor}

First and second generation (weak scale) universality is built into the pMSSM in order to impose a degeneracy solution to the SUSY flavor problem,
while CP violating phases are ignored. The flavor degeneracy solution 
is generally {\it not} to be expected in gravity-mediation models.
One may invoke GMSB or inoMSB or AMSB or CMSSM for flavor universality, but these models are disfavored by $m_h\simeq 125$ GeV and 
naturalness\cite{Baer:2014ica}.
Gravity-mediation can be consistent with flavor and CP constraints by
invoking instead the decoupling solution via 10-100 TeV matter scalars
with the lower range working if there is quasi-degeneracy to within 
tens-of-percent in first/second generation masses. The latter situation is indeed favored by statistical scans of the string 
landscape\cite{Baer:2017uvn}. 

\subsection{pMSSM scan limits and parameter scans}
\label{ssec:scanlimits}

Another issue pertaining to pMSSM scans involves the manner in which parameters are scanned, and the (ad-hoc) upper limits on scan parameters.
For instance, in the Snowmass scans\cite{Dickinson:2022hus}, gaugino and higgsino masses are scanned up to 25 TeV. Slepton masses are scanned up to 25 TeV while squark masses to 50 TeV. Thus, the bulk of 
scanned parameter space is highly unnatural under $\Delta_{EW}$. 
Furthermore, some parameters are scanned via log sampling (which favors lower soft term values) while others are scanned linearly. 
Other scans may use a Markov chain sampling routine\cite{vanBeekveld:2016hug}.
Results displayed in terms of number of surviving models over total number of
sampled models will clearly depend on the sampling details.

\subsection{Dark matter constraints}
\label{ssec:DM}

Some pMSSM scans include various WIMP dark matter constraints including 1. obtaining the measured relic density of DM, 2. obtaining less than or equal to the correct relic density (mixed dark matter but with a standard underabundance of WIMPs or 3. appropriate direct and/or indirect WIMP detection rates provided SUSY neutralinos make up the entire DM abundance.

Basing conclusions on these constraints can be risky business. 
Gelmini and Gondolo emphasized long ago\cite{Gelmini:2006pw} that one might get any value of relic abundance $\Omega_{\chi}h^2$ from SUSY WIMPs when the model is augmented with scalar fields $\phi$, 
representative of stringy moduli. 
If the PQ solution to the strong CP problem is invoked, then one expects also axion dark matter along with axinos and saxions\cite{Baer:2011hx}. The presence of these latter three in the early universe can either augment or detract from the standard abundance\cite{Bae:2014rfa} (labeled here as $\Omega_{\chi}^{std}h^2$). Gravitinos may also add  to the standard relic density\cite{Nakamura:2006uc}. And stringy moduli fields $\phi_i$ can engage in late-time decays (after neutralino freeze-out) to augment the relic density or provide entropy dilution which decreases the expected abundance\cite{Baer:2023bbn}. 

In addition, recently it has been 
noted\cite{Baer:2025oid,Baer:2025srs} that in models where the 
SUSY $\mu$ problem is solved via discrete 
$R$-symmetries\cite{Lee:2011dya}, and where a global $U(1)_{PQ}$ emerges automatically to solve the strong CP problem, then small $R$-parity violation can be expected 
at levels of $(f_a/m_P)^n$. For the cases where $n=1$, then all WIMPs
would decay in the early universe before the onset of BBN, leaving all axion and no WIMP dark matter.

\section{pMSSM scans}
\label{sec:scan}

In this Section, we show results from pMSSM scans that illustrate some of
the issues brought forth in the previous Section.

\subsection{Scan {\bf S}}

We adopt the Isajet 7.91 program\cite{Paige:2003mg} to scan over the 
pMSSM19 parameter space, where our parameter scan limits are taken to coincide  (roughly) with
the {\it Snowmass} scan (Scan {\bf S}) of Ref. \cite{Dickinson:2022hus}.
We assume the elements of the 16-dimensional
spinor rep of $SO(10)$ are all independent $m_{Q_i}\ne m_{U_i}\ne m_{D_i}\ne m_{L_i}\ne m_{E_i}$
for each generation $i=1-3$. 
We further take the first two generations as degenerate $m_j(1)=m_j(2)$
(where the index $j$ runs over the five matter scalars in each generation).
We adopt the Isajet soft term input scale as $Q=51$ TeV, just above the largest of the soft term limits, and then scan over:
\bi
\item $m_Q(1,2),\ m_U(1,2),\ m_D(1,2),\ m_L(1,2),\ m_E(1,2): 0.1-25$ TeV,
\item $m_Q(3),\ m_U(3),\ m_D(3),\ m_L(3),\ m_E(3): 0.1-25$ TeV,
\item $|M_1|: 0.01-25$ TeV,
\item $|M_2|: 0.07-25$ TeV, 
\item $|M_3|: 0.2-50$ TeV, 
\item $|A_k|: 0.001-7$ TeV\ \ ($k=t,b,\tau$), 
\item $|\mu |: 0.1-0.35$ TeV, 
\item $m_A: 0.1-25$ TeV, 
\item $\tan\beta : 1-60$. 
\ei

We further require that the lightest neutralino $\tchi_1^0$ is the LSP,
that $122\ {\rm GeV}<m_h<128$ GeV and that $m_{\tchi_1^+}>0.1$ TeV (LEP2).
For the $\mu$ vs $m^0$ plots (higgsino discovery plane), we also require that $m_{\tg}>2.2$ TeV and $m_{\tst_1}>1.2$ TeV.

\subsubsection{Naturalness from scan {\bf S}}

A difficulty with the Snowmass pMSSM scans was that no natural SUSY points 
with $\Delta_{EW}\alt 30$ could be found in spite of the rather extensive scan. This is understandable from the large scan range of soft terms in
scan {\bf S} where almost always some 3rd generation squark soft terms will lie in the $\agt 10$ TeV range, and gluinos are likely to be extremely heavy as well.  

\begin{figure}[htb!]
\centering
    {\includegraphics[height=0.4\textheight]{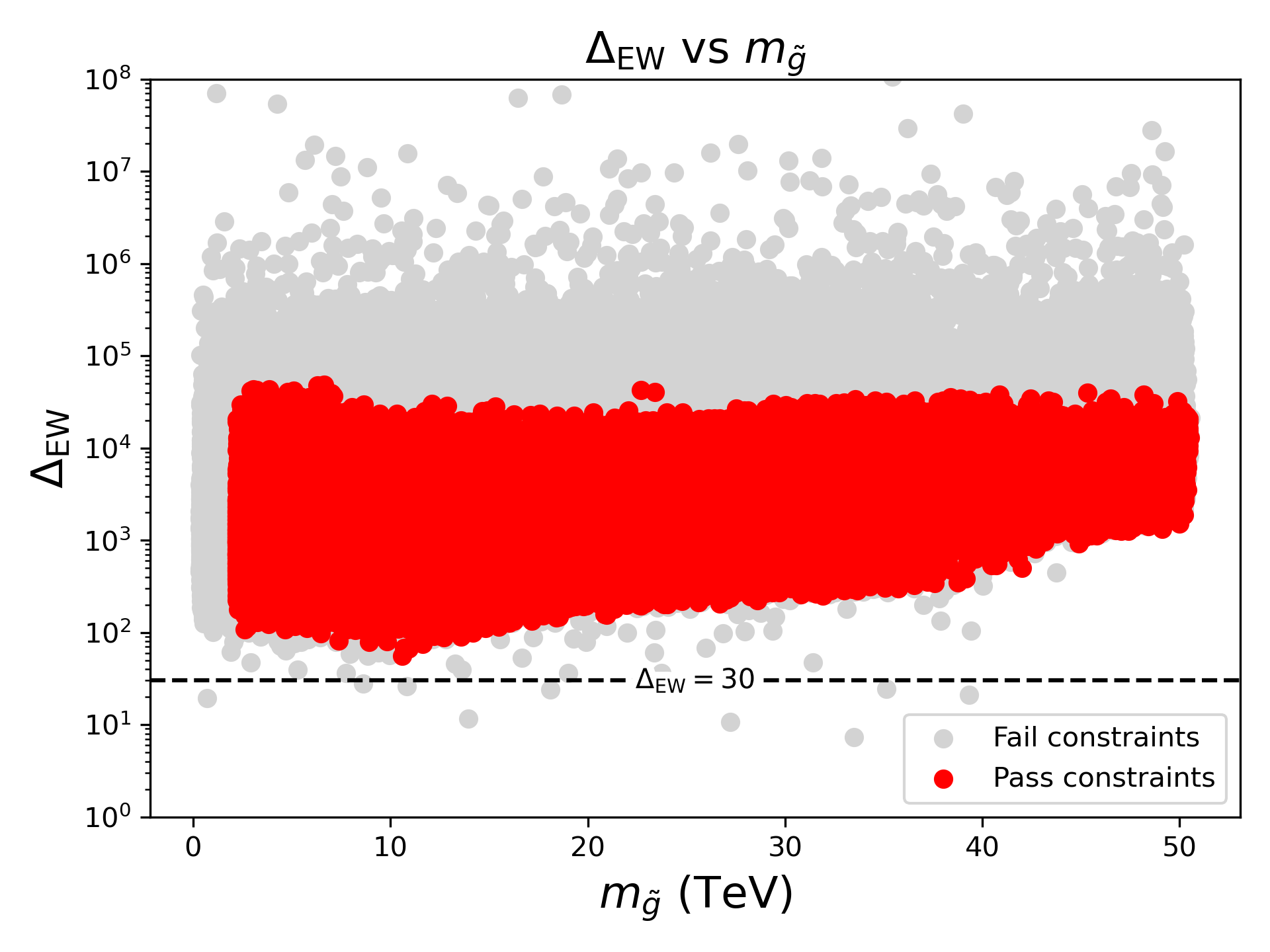}}\\
        {\includegraphics[height=0.4\textheight]{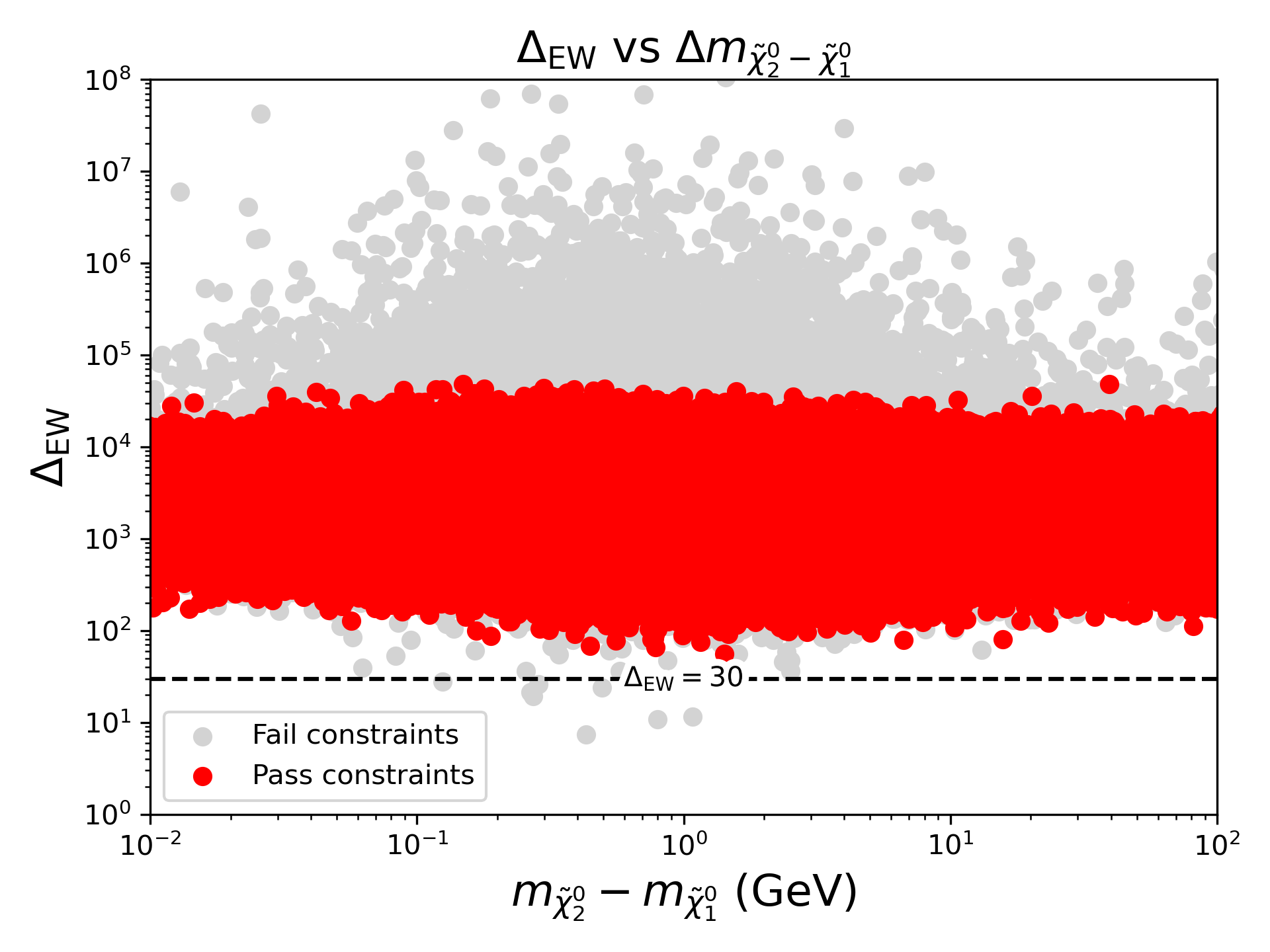}}
        \caption{Plot of {\it a}) $\Delta_{EW}$ vs $m_{\tg}$ and
        {\it b}) $\Delta_{EW}$ vs. $m^0$ from pMSSM19 scan {\bf S} 
        with $10^6$ points.
      \label{fig:dew}}
\end{figure}

We show the value of $\Delta_{EW}$ vs. {\it a}) $m_{\tg}$ and {\it b}) vs. $\Delta m^0\equiv m_{\tchi_2^0}-m_{\tchi_1^0}$ in Fig. \ref{fig:dew}. 
From frame {\it a}), the red points pass the above constraints. 
But in spite of scanning $10^6$ points, no viable ones reach lower than 
$\Delta_{EW}\sim 50$. The lowest $\Delta_{EW}$ points occur for 
$m_{\tg}\sim 10$ TeV but as $m_{\tg}$ increases, the surviving points have
increased finetuning. 
In frame {\it b}), the naturalness measure 
reaches a minimum for mass gap $\Delta m^0\equiv m_{\tchi_2^0}-m_{\tchi_1^0}\sim 1-2$ GeV and becomes larger for smaller 
mass gaps, and modestly larger for bigger mass gaps. 
This raises concerns that LHC searches for nearly pure higgsinos with long lifetimes and tiny mass gaps are actually probing a rather implausible, unnatural region of parameter space which requires a form of gaugino masses
$M_1,M_2\gg M_3$ which is hard to realize in theoretically motivated models.

In Fig. \ref{fig:plane1} frame {\it a}), we show scan {\bf S} in the
$m_{\tq}$ vs. $m_{\tg}$ plane, where $m_{\tq}$ is the average over all
squark masses. 
Very low $m_{\tq}$ mass values lead to CCB minima as do 
very heavy squark masses (due to slight 2-loop RG running effects below $Q=51$ TeV). 
From the plot, we see only a few points with 
$\Delta_{EW}<100$ and no points with $\Delta_{EW}<50$. 
The ATLAS mass bounds from the $m_{\tq}$ vs. $m_{\tg}$ plane are shown as
red dots in the lower-left side of the plot. 
From this plot, using the huge scan {\bf S} upper parameter limits,
one might conclude that LHC excludes only a tiny portion of allowed 
pMSSM19 parameter space, and that SUSY is almost never natural. 
In Fig. \ref{fig:plane2}{\it a}) we
show the same scan {\bf S} points (with the added gluino and stop simplified model bounds) in the $m_{\tchi_1^0}$ vs. $\Delta m^0$ plane. The allowed points reach down to sub-GeV mass gaps but only a handful semi-natural points lie in this region.

If we implement gaugino mass universality so that weak scale gauginos
occur in the ratio $M_3:M_2:M_1\sim 7:2:1$ then we arrive at Fig. \ref{fig:plane1}{\it b}) which we call scan ${\bf S}_u$ ($u$ for unified gaugino masses). 
In this case, it is somewhat easier to generate natural points since for low $M_3$ values, then $M_1$ and $M_2$ are bounded from becoming too large. 

In Fig. {\ref{fig:plane2}}{\it b}), the scan ${\bf S}_u$ is shown in the
higgsino discovery plane. In this case, the lower portion of the discovery plane with sub-GeV neutralino mass splitting is cut off because points
with $M_1,M_2\gg M_3$ cannot be generated. The semi-natural points 
with $\Delta_{EW}<100$ all have mass gaps $\Delta m^0\agt 3$ GeV.

\subsection{Scan {\bf T}}

To gain better accord with naturalness, we compare scan {\bf S} with a
{\it targeted} parameter space scan {\bf T} that should be more
focused in picking out natural SUSY solutions since it avoids 
parameter values in the tens-of-TeV regime which are almost always  unnatural. 
The input scale is also at $Q=51$ TeV as in scan {\bf S}.
\bi
\item $m_Q(1,2),\ m_U(1,2),\ m_D(1,2),\ m_L(1,2),\ m_E(1,2): 0.5-25$ TeV,
\item $m_Q(3),\ m_U(3),\ m_D(3),\ m_L(3),\ m_E(3): 0.1-10$ TeV,
\item $|M_1|: 0.01-5$ TeV,
\item $|M_2|: 0.07-5$ TeV, 
\item $|M_3|: 0.2-5$ TeV, 
\item $|A_k|: (0.001-2)m_0(3)$ TeV, 
\item $|\mu |: 0.1-0.35$ TeV, 
\item $m_A: 0.1-10$ TeV, 
\item $\tan\beta : 3-60$. 
\ei

From Fig. \ref{fig:plane1}{\it c}), we see the $m_{\tq}$ vs. $m_{\tg}$ plane
has an (artificial) upper bound $m_{\tg}\alt 5$ TeV due to the new scan limits. Also, it is much easier to generate natural SUSY p-space points, even some with $\Delta_{EW}\alt 30$. For this scan, one might conclude that ATLAS/CMS exclude about half of p-space, and also that natural SUSY is relatively common. From Fig. \ref{fig:plane2}{\it c}), the red low $\Delta_{EW}$ points nearly disappear due to LHC simplified model mass bounds and some semi-natural points have sub-GeV mass gaps. But if we proceed to scan ${\bf T}_u$ in
Fig. \ref{fig:plane1}{\it d}), we obtain fractionally many more natural points, but from Fig. \ref{fig:plane2}{\it d}), they all occur with  higgsino mass gaps $\agt 5$ GeV. 

\subsection{Scan {\bf U}}

The pMSSM tries to be general in implementing 19 free parameters
at the weak scale. But the generality of 19 free parameters can be maintained
without discarding the successes of SUSY RG evolution
(gauge coupling evolution, radiatively-driven EW symmetry breaking via the
large top quark Yukawa coupling) by implementing the 19 free parameters
instead at the {\it unification} scale $Q\simeq 2\times 10^{16}$ GeV,
which we call scan {\bf U}.

For scan {\bf U}, we implement the 9-free-GUT-scale parameter scan (SUGRA9) of
\bi
\item $m_0(1,2): 10-40$ TeV,
\item $m_0(3): 0.1-10$ TeV,
\item $|M_1|: 0.1-10$ TeV,
\item $|M_2|: 0.1-10$ TeV,
\item $M_3: 0.1-10$ TeV,
\item $A_0: m_0(3)- 3m_0(3)$, 
\item $\mu (weak): 0.1-.35$ TeV,
\item $m_A: 0.3-10$ TeV and
\item $\tan\beta :3-60$. 
\ei

\begin{figure}[htb!]
\centering
    {\includegraphics[height=0.2\textheight]{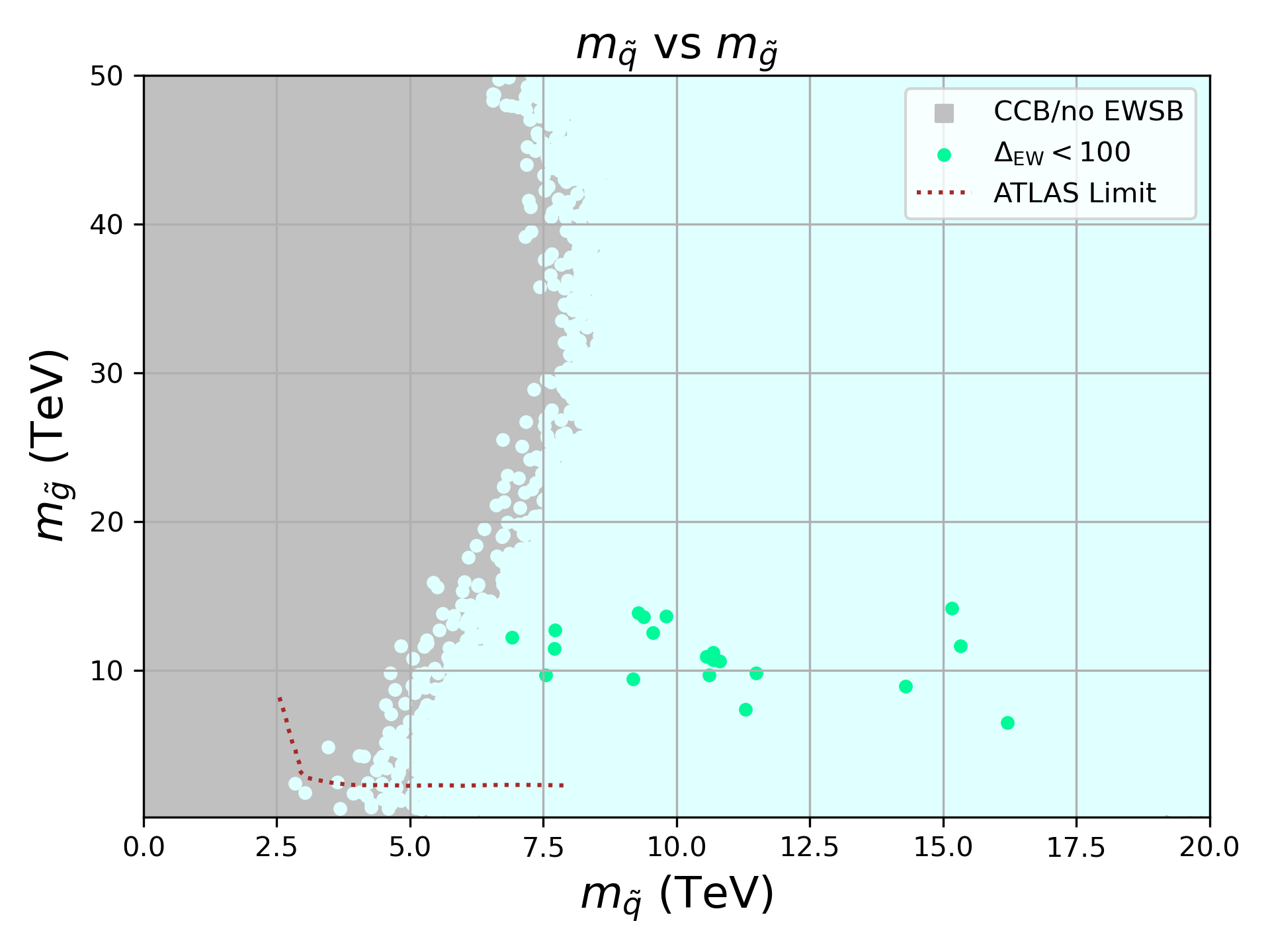}}
        {\includegraphics[height=0.2\textheight]{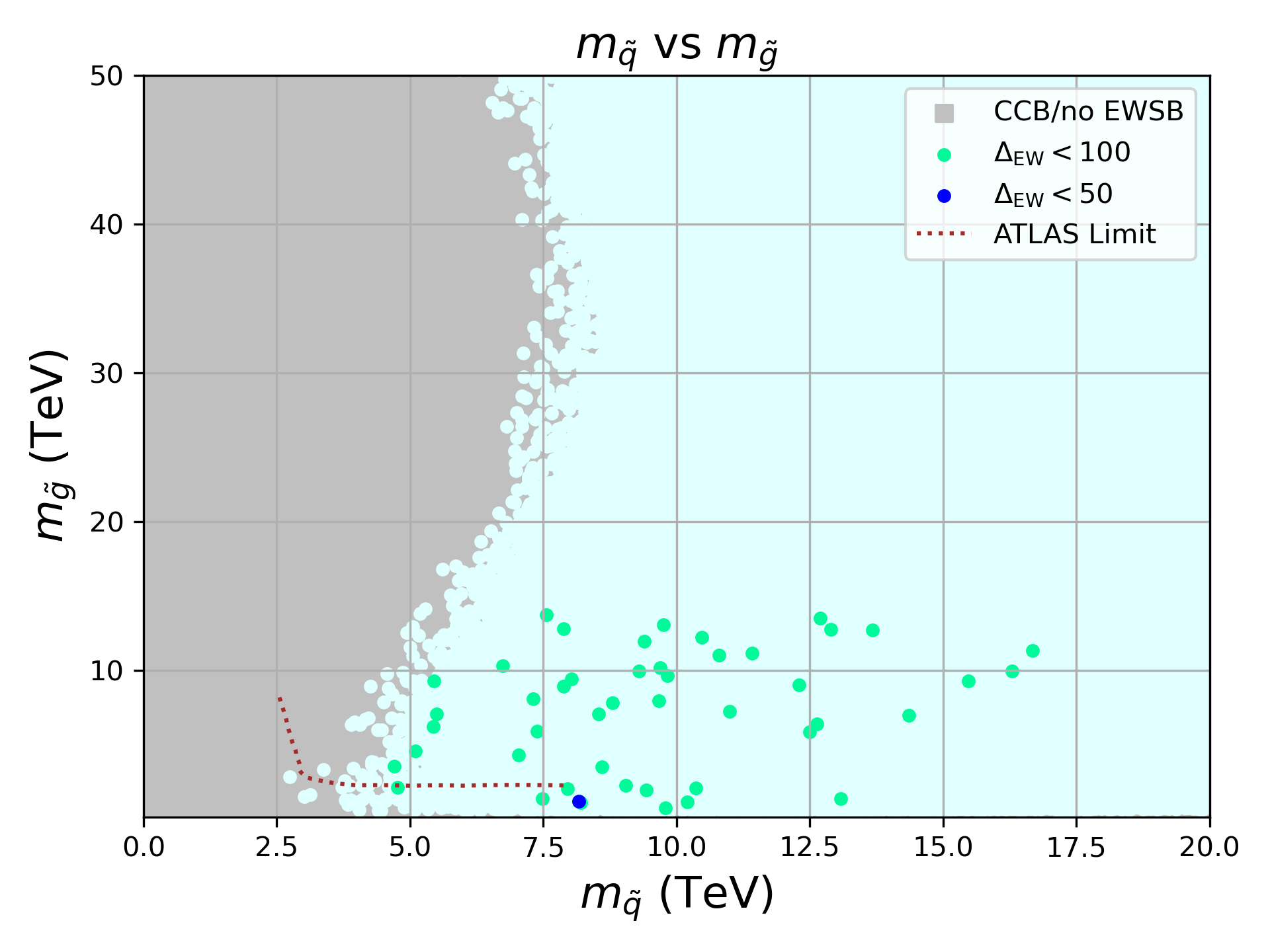}}\\
        {\includegraphics[height=0.2\textheight]{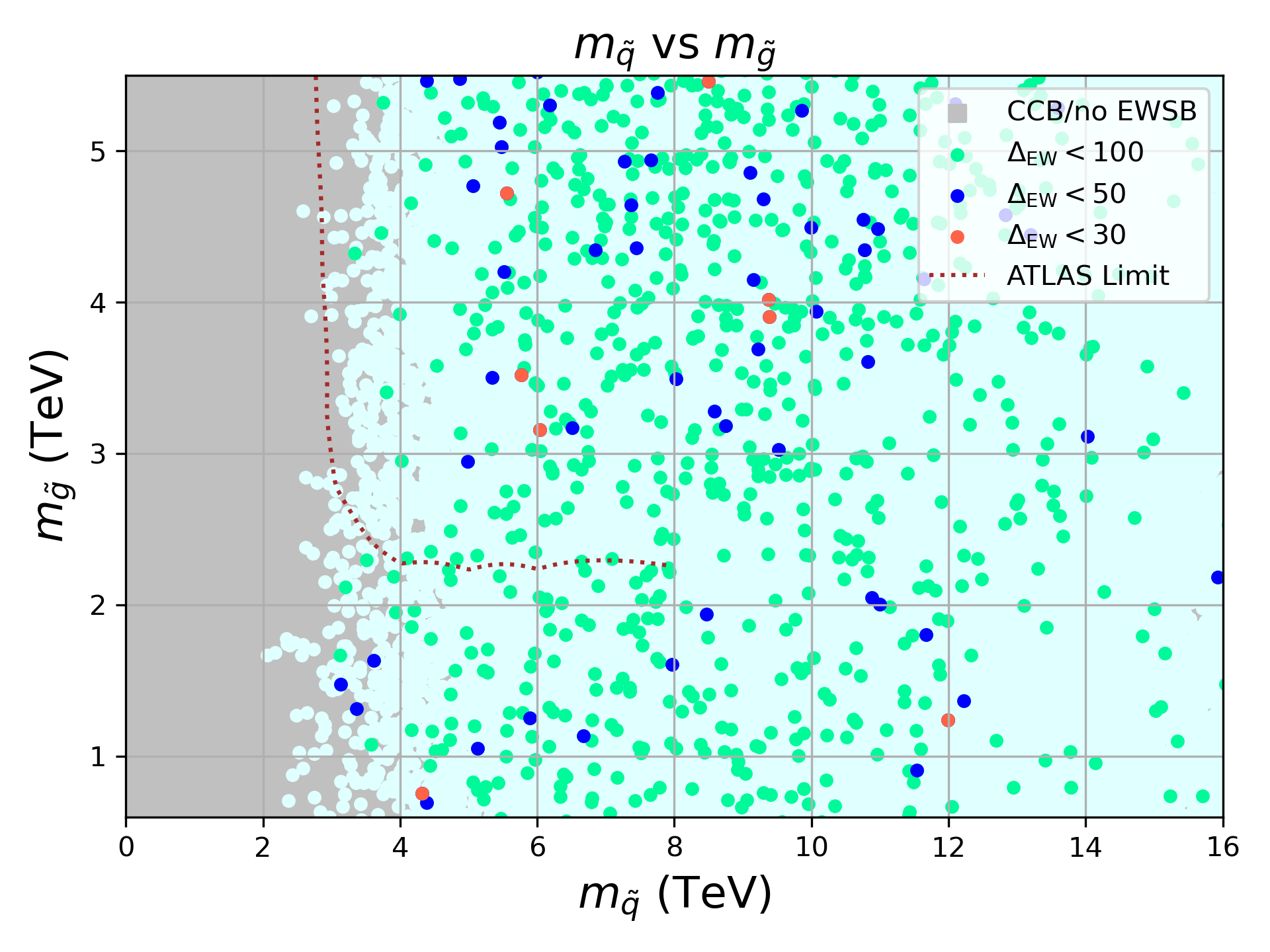}}
                {\includegraphics[height=0.2\textheight]{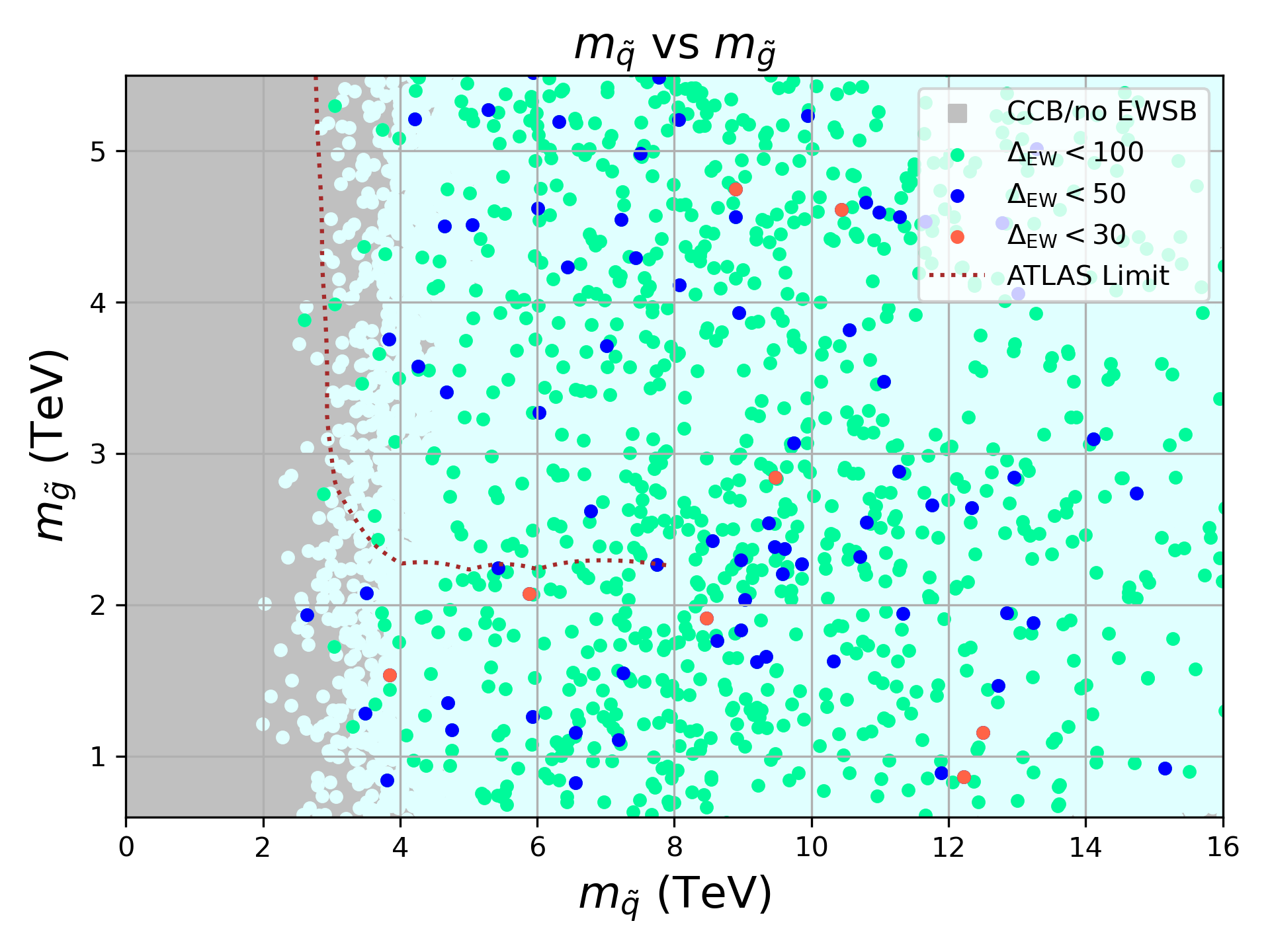}}\\
        {\includegraphics[height=0.2\textheight]{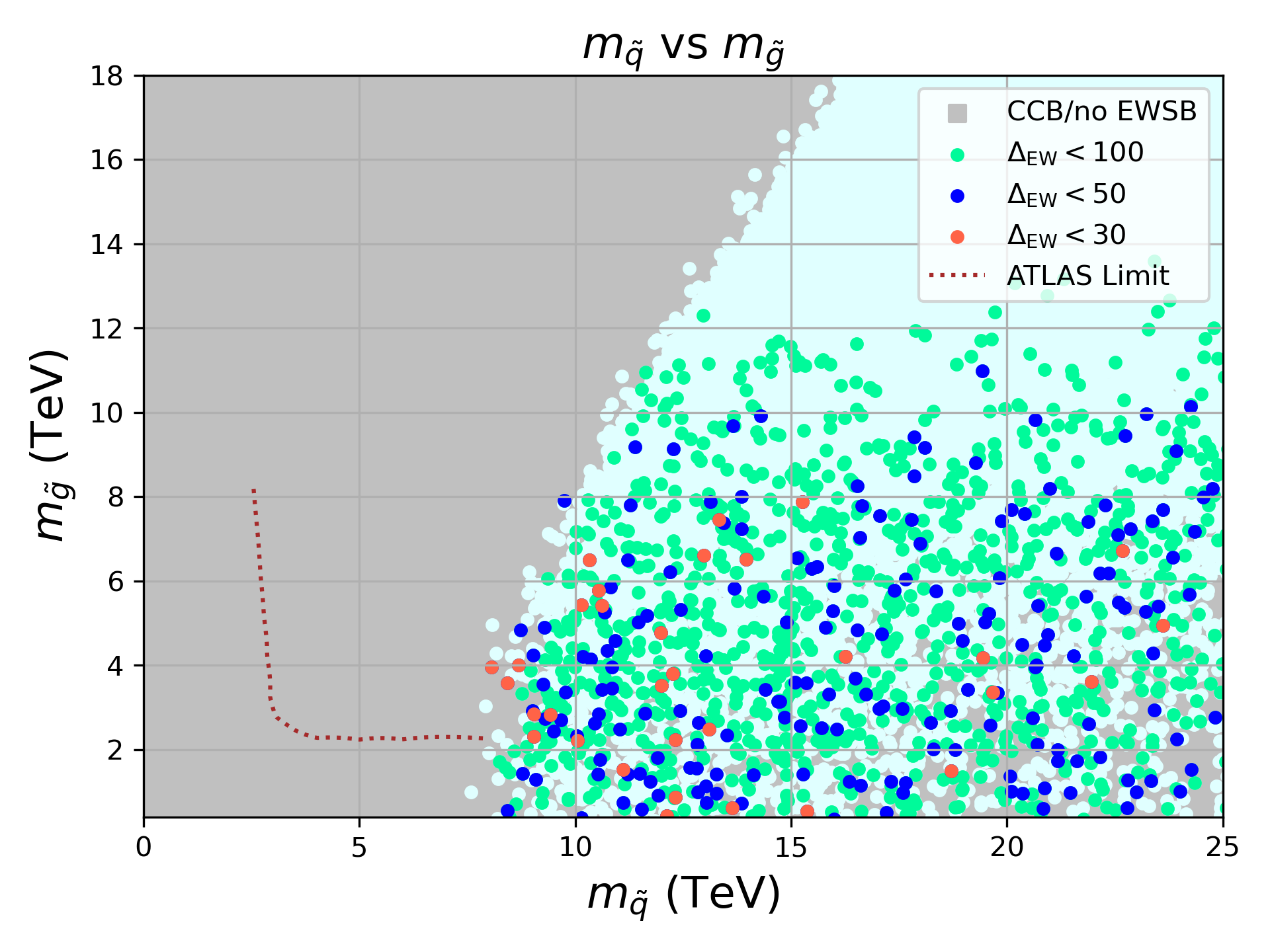}}
                {\includegraphics[height=0.2\textheight]{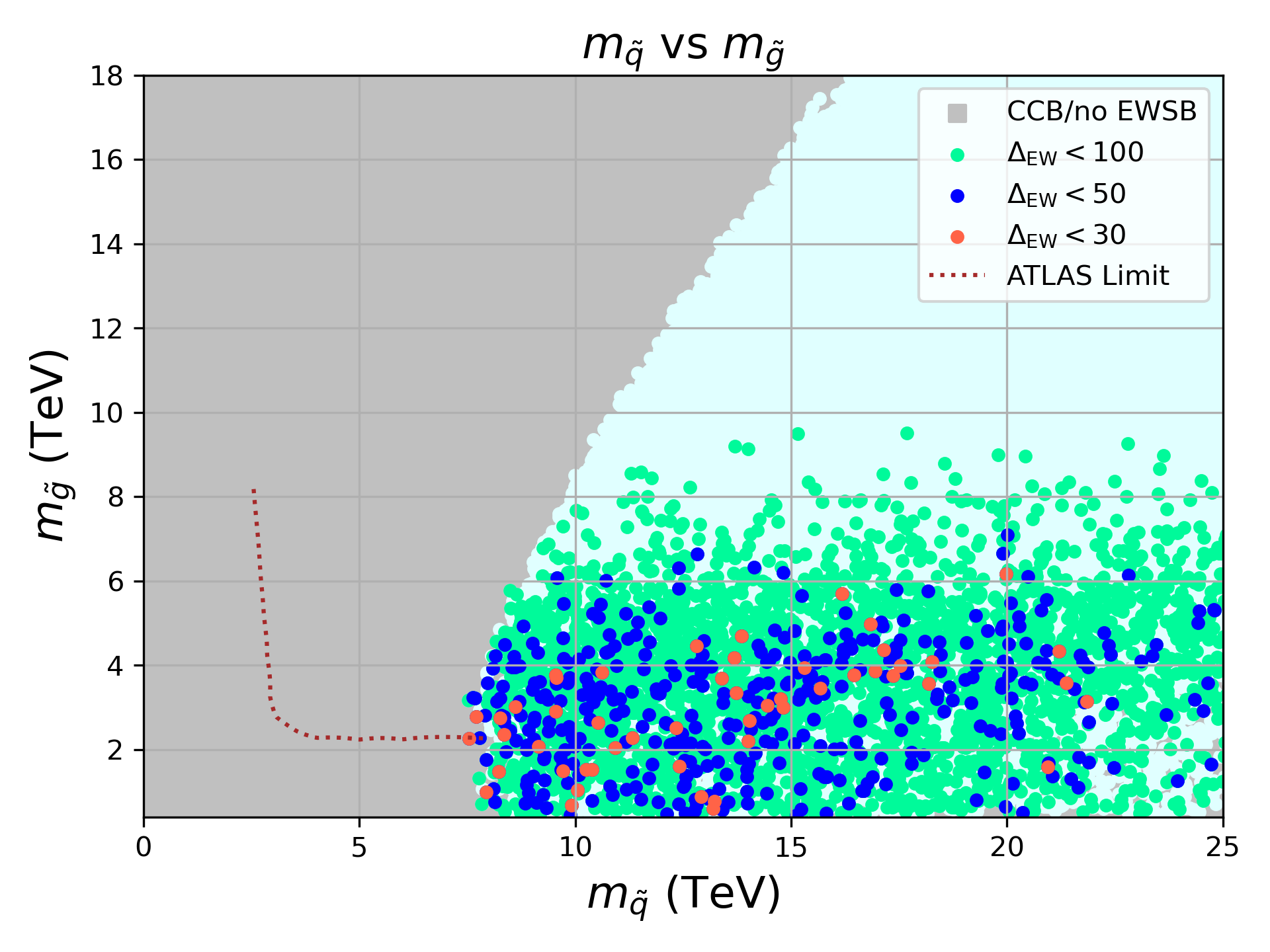}}
        \caption{Plot of pMSSM scan points in the $m_{\tq}$ vs. $m_{\tg}$
          plane for {\it a}) scan {\bf S}, {\it b}) scan ${\bf S_u}$ 
          with gaugino mass unification $M_3=3.5M_2=7M_1$, {\it c}) scan {\bf T} and
          {\it d}) scan ${\bf T_u}$ with gaugino mass unification.
          In frame {\it e}), we plot results from the SUGRA9 model
          (scan {\bf U}) and in {\it f}), we plot the scan ${\bf U_u}$ results
          with gaugino mass unification $M_1=M_2=M_3\equiv m_{1/2}$
          at the GUT scale.
      \label{fig:plane1}}
\end{figure}

\begin{figure}[htb!]
\centering
    {\includegraphics[height=0.2\textheight]{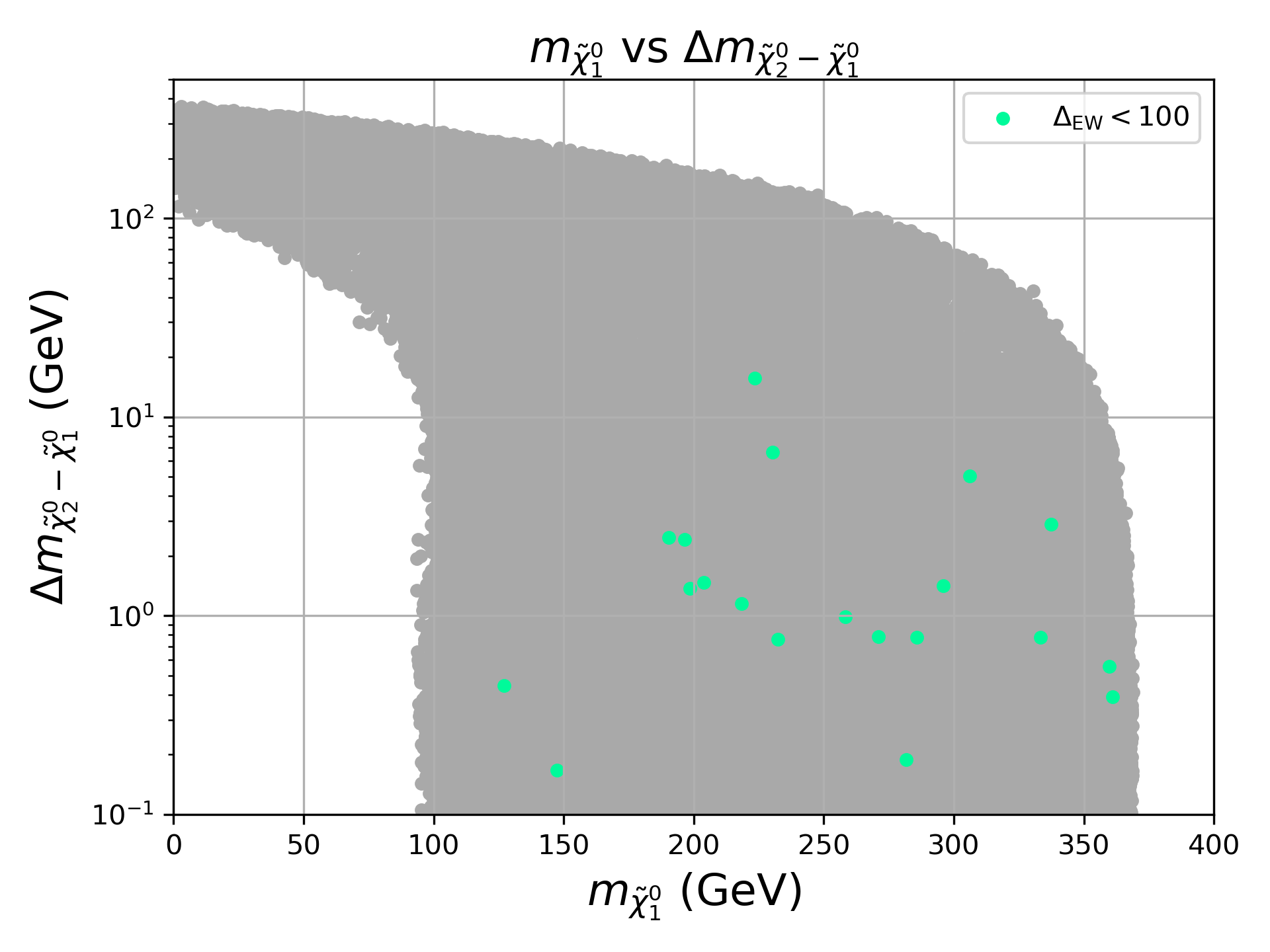}}
        {\includegraphics[height=0.2\textheight]{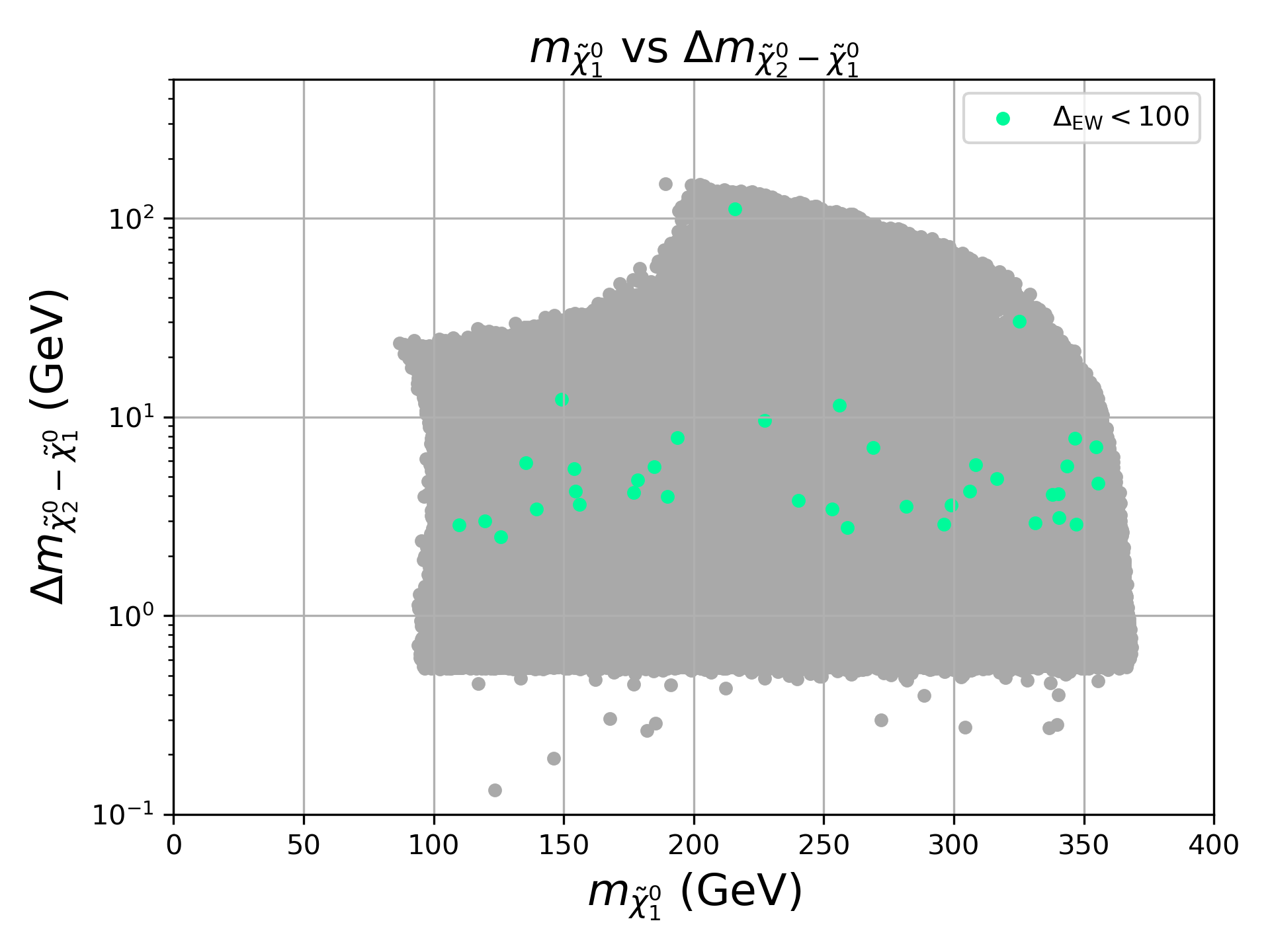}}\\
        {\includegraphics[height=0.2\textheight]{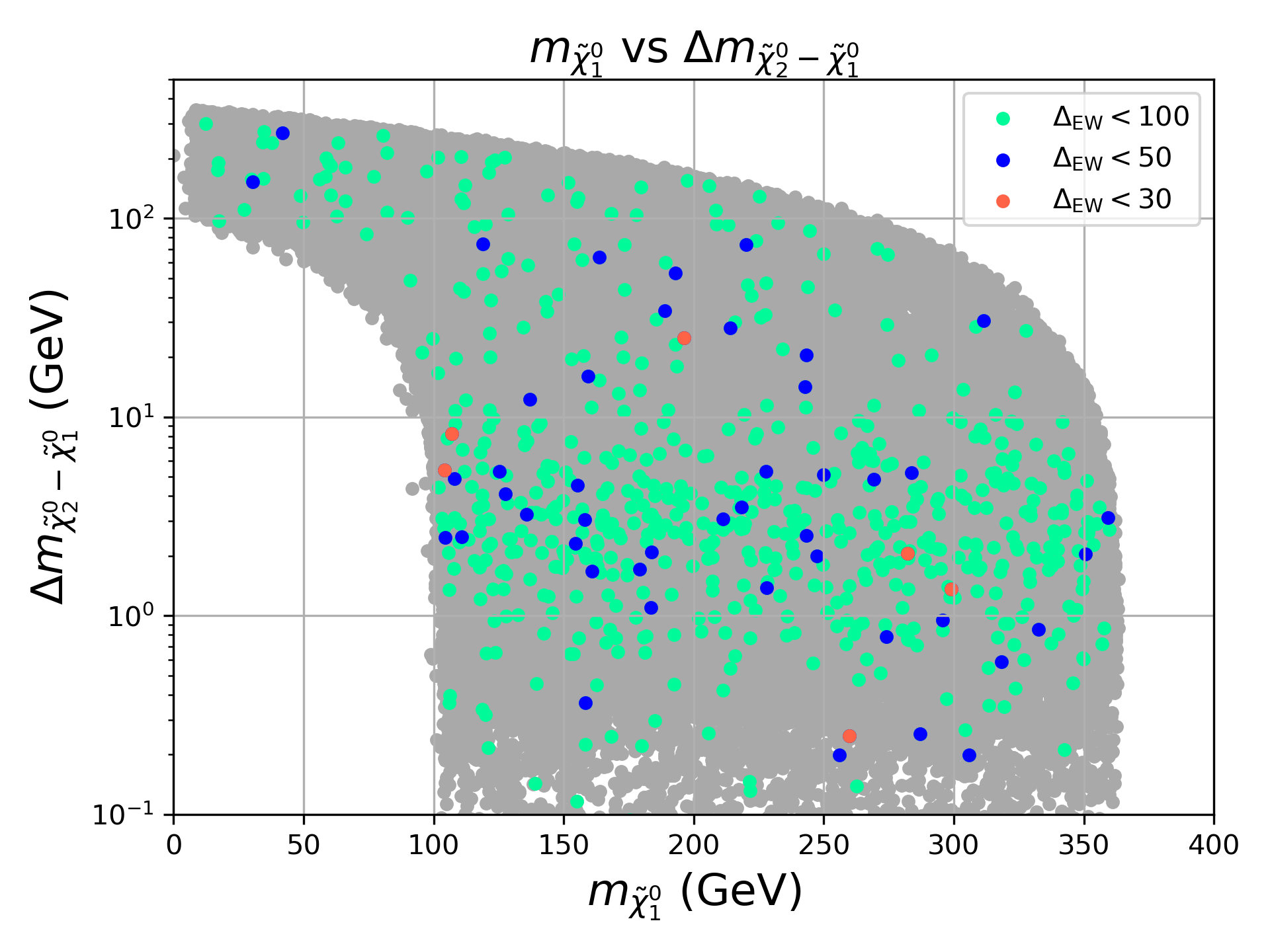}}
                {\includegraphics[height=0.2\textheight]{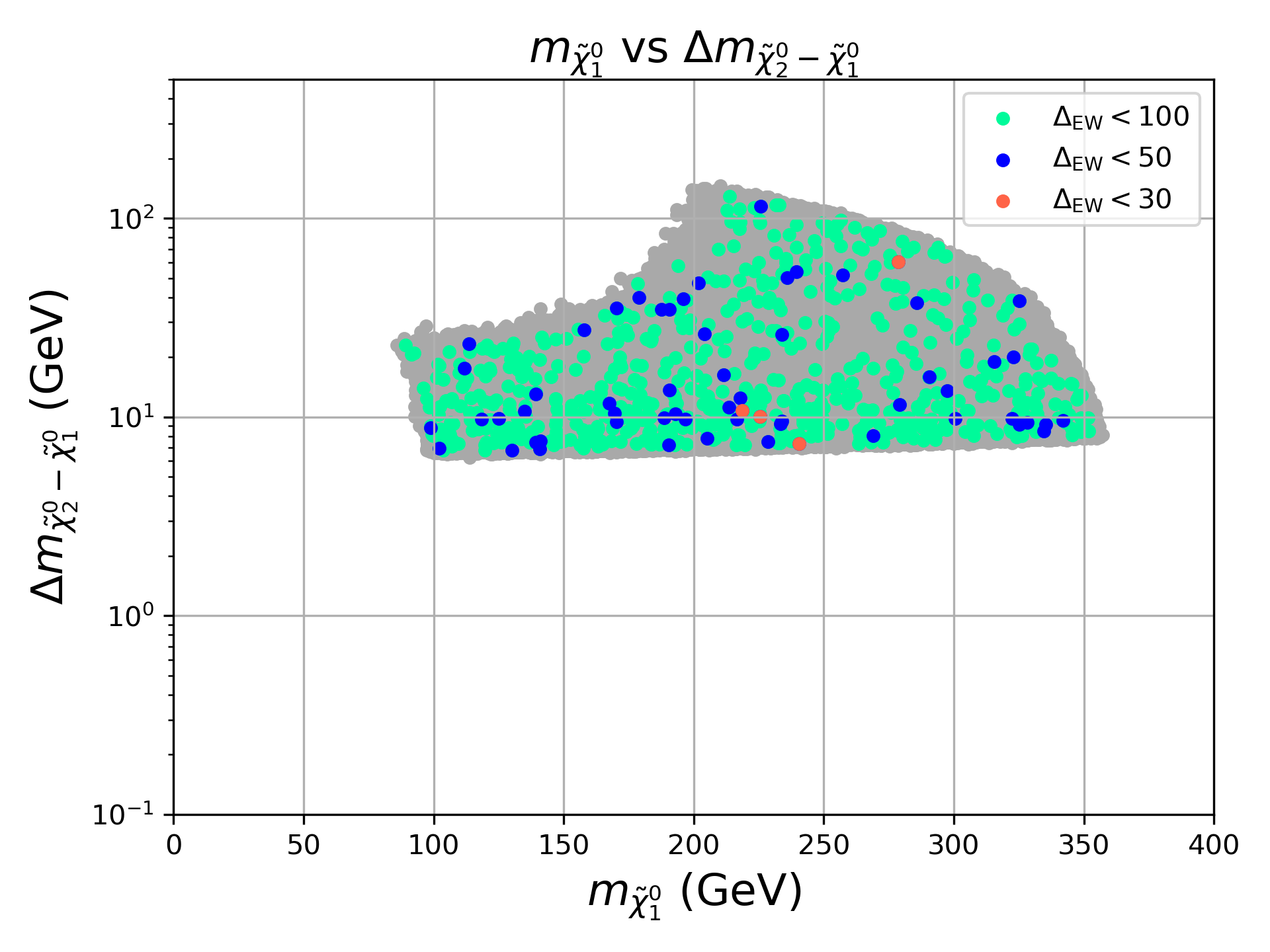}}\\
        {\includegraphics[height=0.2\textheight]{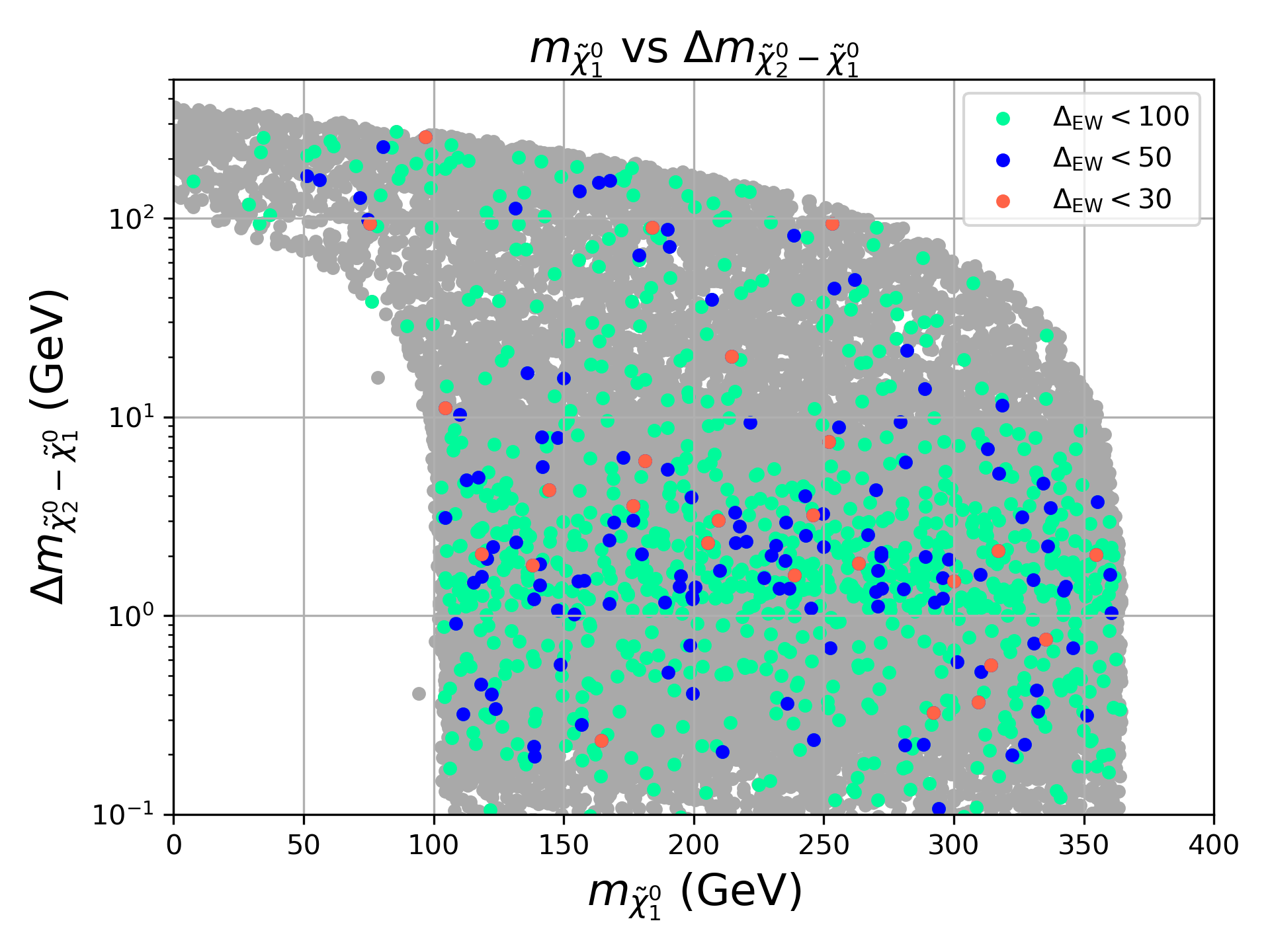}}
                {\includegraphics[height=0.2\textheight]{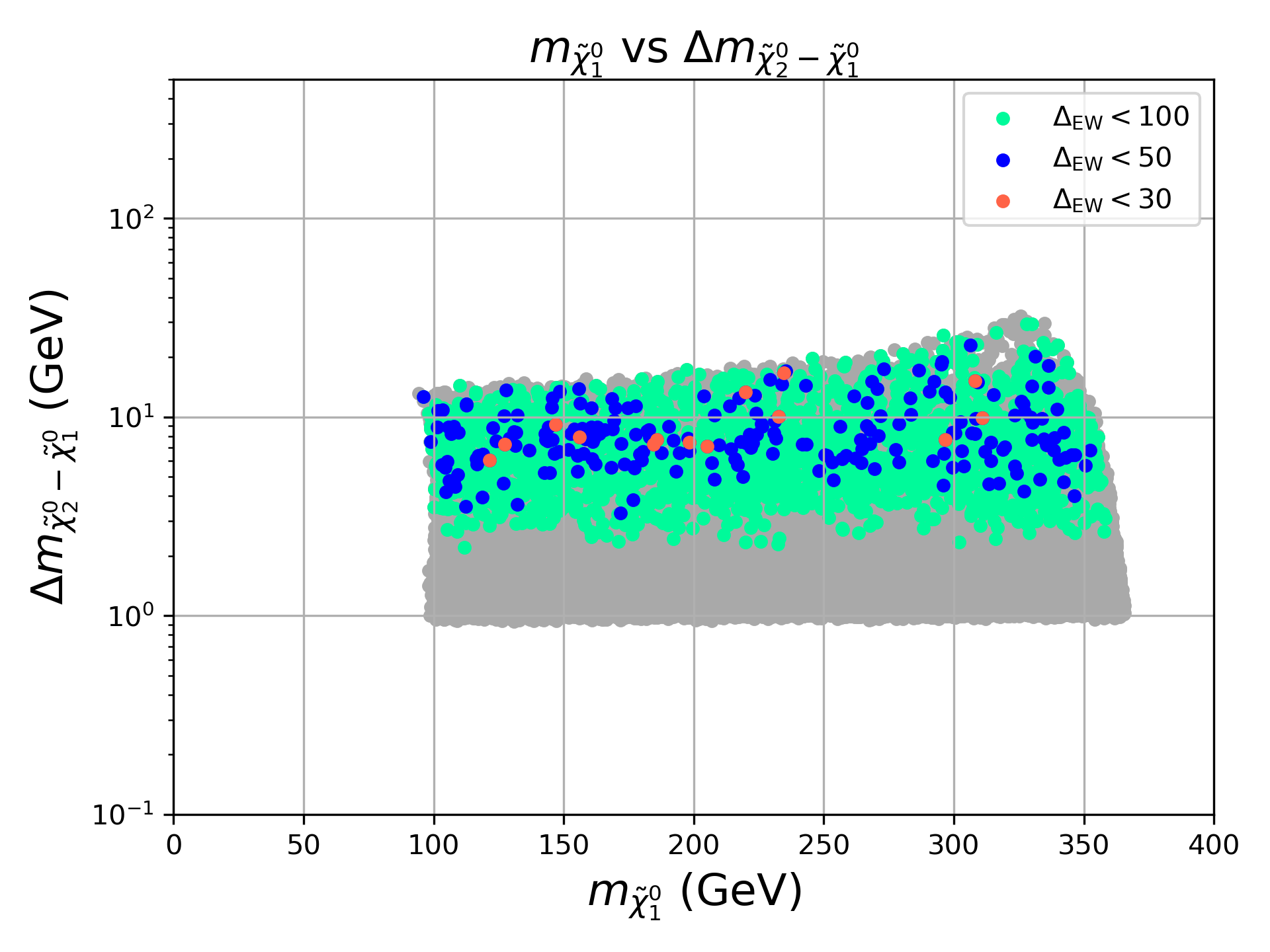}}
        \caption{Plot of pMSSM scan points in the $\mu$ vs. $\Delta m^0$
          plane for {\it a}) scan {\bf S}, {\it b}) scan ${\bf S_u}$ with
          gaugino mass unification $M_3=3.5M_2=7M_1$, {\it c}) scan {\bf T} and
          {\it d}) scan ${\bf T_u}$ with gaugino mass unification.
          In frame {\it e}), we plot results from the SUGRA9 model
          (scan {\bf U}) and in {\it f}) we plot the scan ${\bf U_u}$ results
          with gaugino mass unification $M_1=M_2=M_3\equiv m_{1/2}$
          at the GUT scale.
      \label{fig:plane2}}
\end{figure}

The results of scan {\bf U} are shown in the gluino vs. squark mass plane in
Fig. \ref{fig:plane1}{\it e}). Due to the large lever-arm of RG running, now a greater portion of parameter space lies in the CCB/noEWSB region and is excluded. The natural SUSY parameter points are now more common but do 
seem to have an upper bound with $m_{\tg}\alt 8-12$ TeV depending on how much finetuning one might tolerate. Scan {\bf U} in the higgsino discovery plane is shown in frame Fig. \ref{fig:plane2}{\it e}) where again 
sub-GeV natural SUSY parameter points can be found due to configurations with $M_1,M_2\gg M_3$. 

If we implement scan {\bf U} with unified GUT-scale gaugino masses $M_1=M_2=M_3\equiv m_{1/2}$ at $Q\simeq 2\times 10^{16}$ GeV, then we arrive at results in Fig. \ref{fig:plane1}{\it f}) (${\bf U}_u$). 
The ${\bf U}_u$ scan finds an even greater proportion of natural SUSY
parameter points with low $\Delta_{EW}$ and an even sharper upper 
bound where $m_{\tg}\alt 6-9$ TeV. 
In Fig. \ref{fig:plane2}{\it f}), we plot scan ${\bf U}_u$ in the higgsino discovery plane and find no viable p-space points with sub-GeV mass gaps 
and no natural SUSY points with mass gaps below $2-5$ GeV, showing the influence of gaugino mass unification. 
The large mass gap solutions are also excluded because these would occur 
for small $M_1,\ M_2\sim \mu$ values while $M_3$ is required to exist beyond
the ATLAS/CMS simplified model gluino mass limits.


%

\section{NUHM4 plane and prospects for LHC via $\tst_1\tst_1^*$ production}
\label{sec:nuhm4}

One thing we have learned from LHC data is that to accommodate 
$m_h\sim 125$ GeV naturally, then one needs large trilinear $A$-terms 
as are generically expected in supergravity models derived from superstrings.
This also favors non-sequestered gaugino masses (gaugino mass unification).
Thus, as a standard supersymmetric model for the mid-term LHC era, gravity-mediation is most favored. 
Hence, instead of the pMSSM, we put forth
the original gravity-mediation (SUGRA) model as the standard. 
We implement the following.
\bi
\item Re-install RG running from the gauge coupling unification scale
$m_{GUT}\sim 2\times 10^{16}$ GeV down to the weak scale.
\item unification of matter scalars of each generation $m_0(i)$ for $i=1-3$
\item otherwise, non-unification of scalar masses in accord with SUGRA expectations: $m_0(1)\ne m_0(2)\ne m_0(3)\ne m_{H_u}\ne m_{H_d}$
\item gaugino mass unification to $m_{1/2}$ at $Q=m_{GUT}$
\item expect the decoupling/quasi-degeneracy solution to the SUGRA flavor and CP problems: $m_0(1)\sim m_0(2)\sim 10-40$ TeV
\ei
This places us in what is called the four-extra-parameter non-universal Higgs model NUHM4 as the low energy effective field theory at $Q<m_{GUT}$ with parameter space 
\be
m_0(1)\sim m_0(2),\ m_0(3),\ m_{1/2},\ A_0,\ \tan\beta,\ \mu,\ m_A
\ee
where the Higgs soft masses have been traded for $\mu$ and $m_A$ at the weak scale\cite{Ellis:2002wv}. (The four extra parameters beyond mSUGRA/cMSSM are
$m_0(2,3)$, $\mu$ and $m_A$.)
We will focus on natural models with low $\Delta_{EW}$ since these give good predictions for the magnitude of the weak scale.

\subsection{Neutralino mass gap in complete models}

\begin{figure}[htb!]
\centering
    {\includegraphics[height=0.4\textheight]{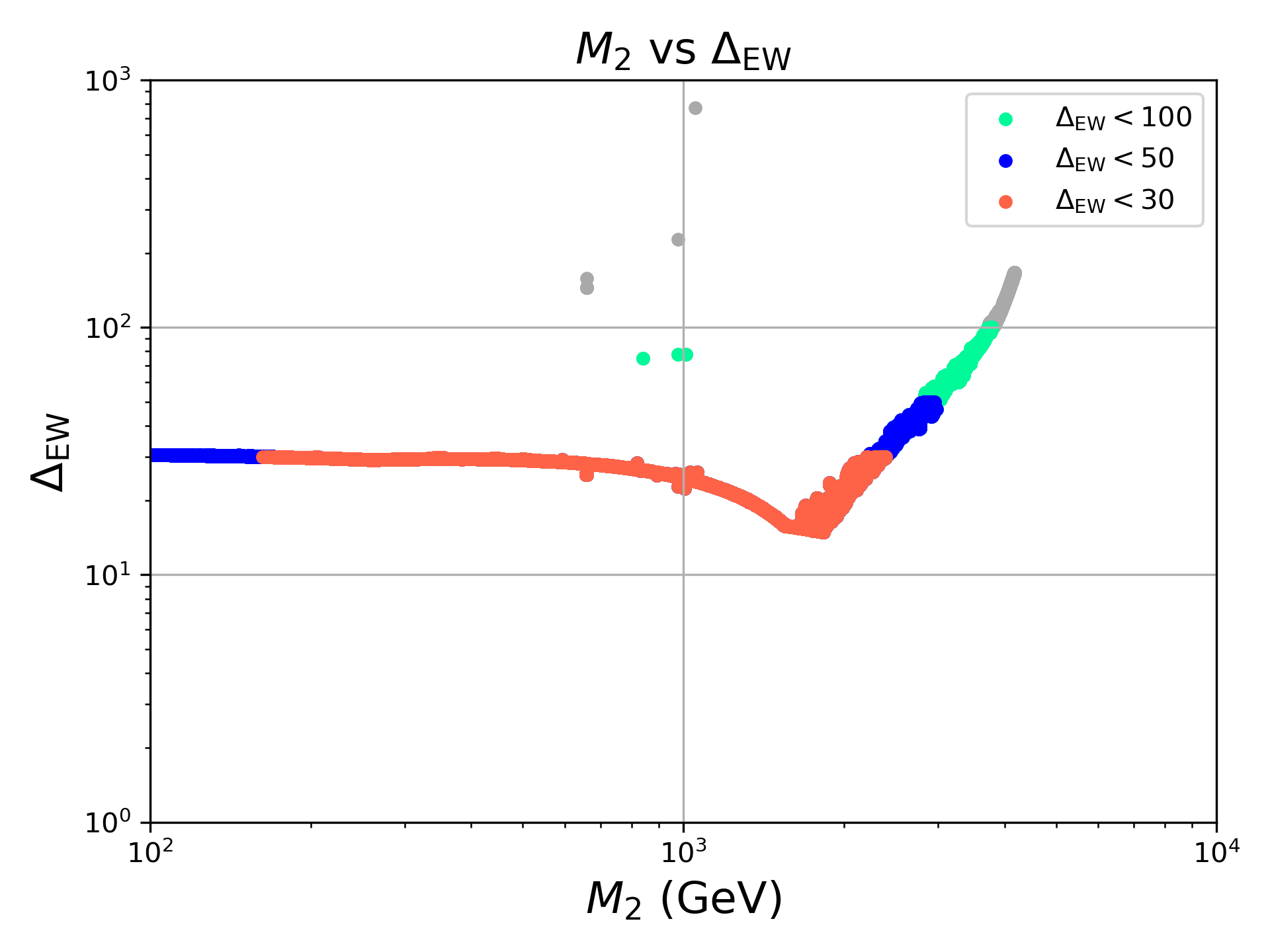}}\\
        {\includegraphics[height=0.4\textheight]{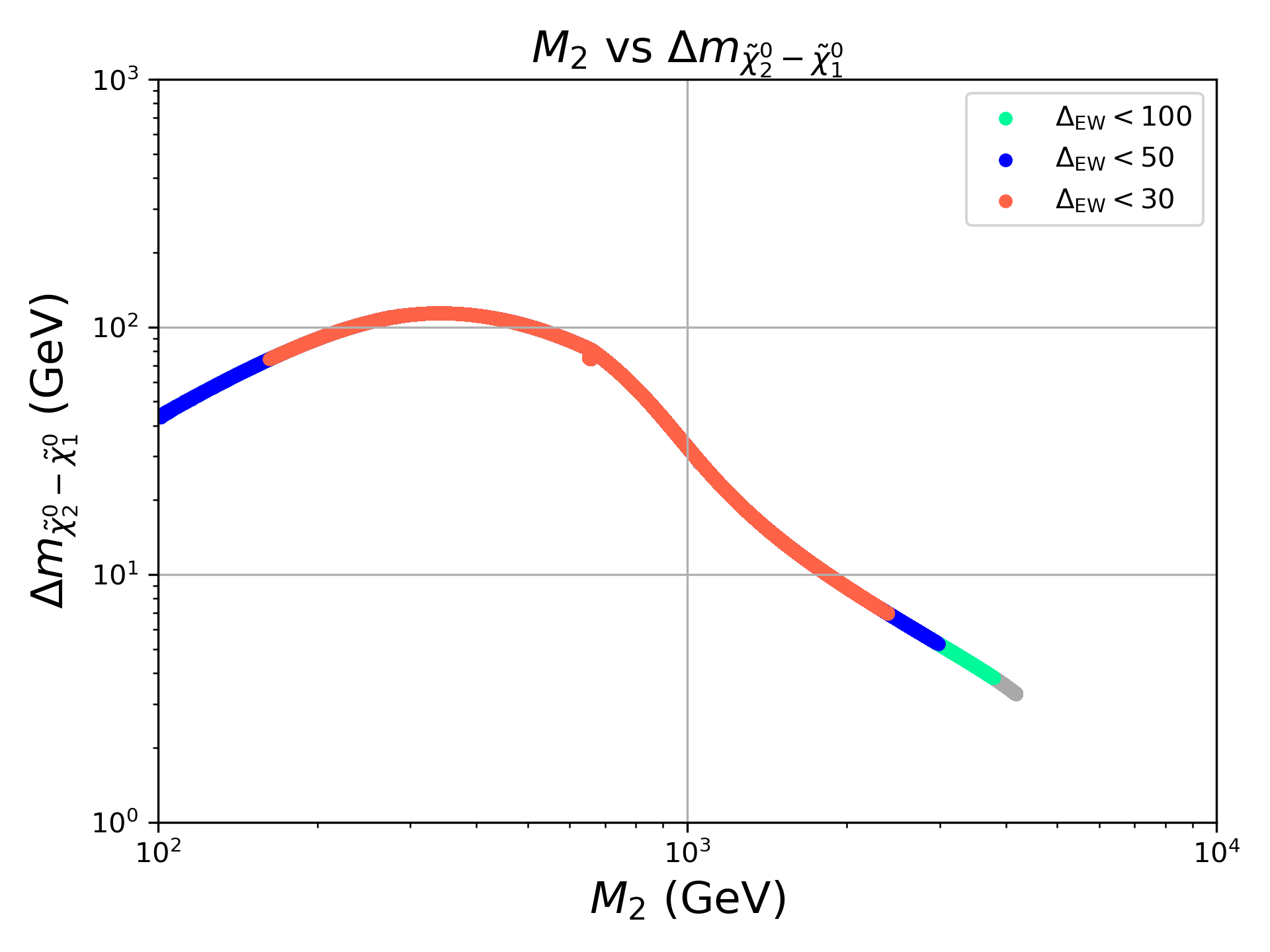}}
        \caption{Plot of {\it a}) $\Delta_{EW}$ and {\it b}) $\Delta m^0\equiv m_{\tchi_2^0}-m_{\tchi_1^0}$ vs. $M_2=2M_1$ 
        for a NUHM4 benchmark point with non-universal gaugino masses
        with $m_0(1,2,3)=5$ TeV, $M_{3}=1.2$ TeV, $A_0=-8$ TeV, $\tan\beta =10$
        with $\mu =200$ GeV and $m_A=2$ TeV.
      \label{fig:M2}}
\end{figure}

In Fig. \ref{fig:M2}, we adopt a NUHM4 benchmark point with
$m_0(1,2,3)=5$ TeV, $m_{1/2}=1.2$ TeV, $A_0=-8$ TeV and $\tan\beta =10$ with
$\mu =200$ GeV and $m_A=2$ TeV. But we take non-unified gaugino masses
with variable $M_2=2M_1$ to show the variation of {\it a}) $\Delta_{EW}$
and {\it b}) $\Delta m^0$ versus $M_2$. From frame {\it a}), we see that for low $M_2$ the value of $\Delta_{EW}$ hardly changes and is fixed 
at nearly $\sim 30$. This is because, since $\mu$ is fixed at 200 GeV, 
the values of $\Sigma_u^u(\tst_{1,2})$ give the biggest contribution to 
$\Delta_{EW}$ and are hardly affected by small $M_2$ or $M_1$. 
As $M_2$ increases, it feeds into $dm_{Q_3}^2/dt$ making the RG running
of $m_{Q_3}^2$ more steep. $\Delta_{EW}$ drops to low values $\alt 20$ 
for $M_2\sim 2$ TeV where cancellations in $\Sigma_u^u(\tst_{1,2})$ 
make the model more natural. As $M_2$ increases even more, the top-squark masses increase as does $\Sigma_u^u$ so that the model becomes unnatural.

In frame {\it b}), we plot the corresponding value of $\Delta m^0$ which
is maximal $\sim 100$ GeV for $M_2\sim 400$ GeV where the lighter neutralinos are mixed bino/wino/higgsino states. For $M_2=2M_1\sim 100$ 
GeV then the mass gap is essentially that of $m(wino)-m(bino)\sim 50$ GeV.
For much higher $M_2\agt 400$ GeV values, then the lightest neutralinos
become increasingly higgsino-like and the mass gap drops to below 10 GeV
for $M_2>m_{1/2}$ and decreases to as low as 4 GeV for $M_2\agt 4$ TeV.
Comparing against frame {\it a}), we see that the smallest mass gaps,
coming from $M_1,M_2\gg m_{1/2}$, correspond to increasingly unnatural models.

\subsection{The $m_0(3)$ vs. $m_{1/2}$ parameter space}

\begin{figure}[htb!]
\centering
    {\includegraphics[height=0.28\textheight]{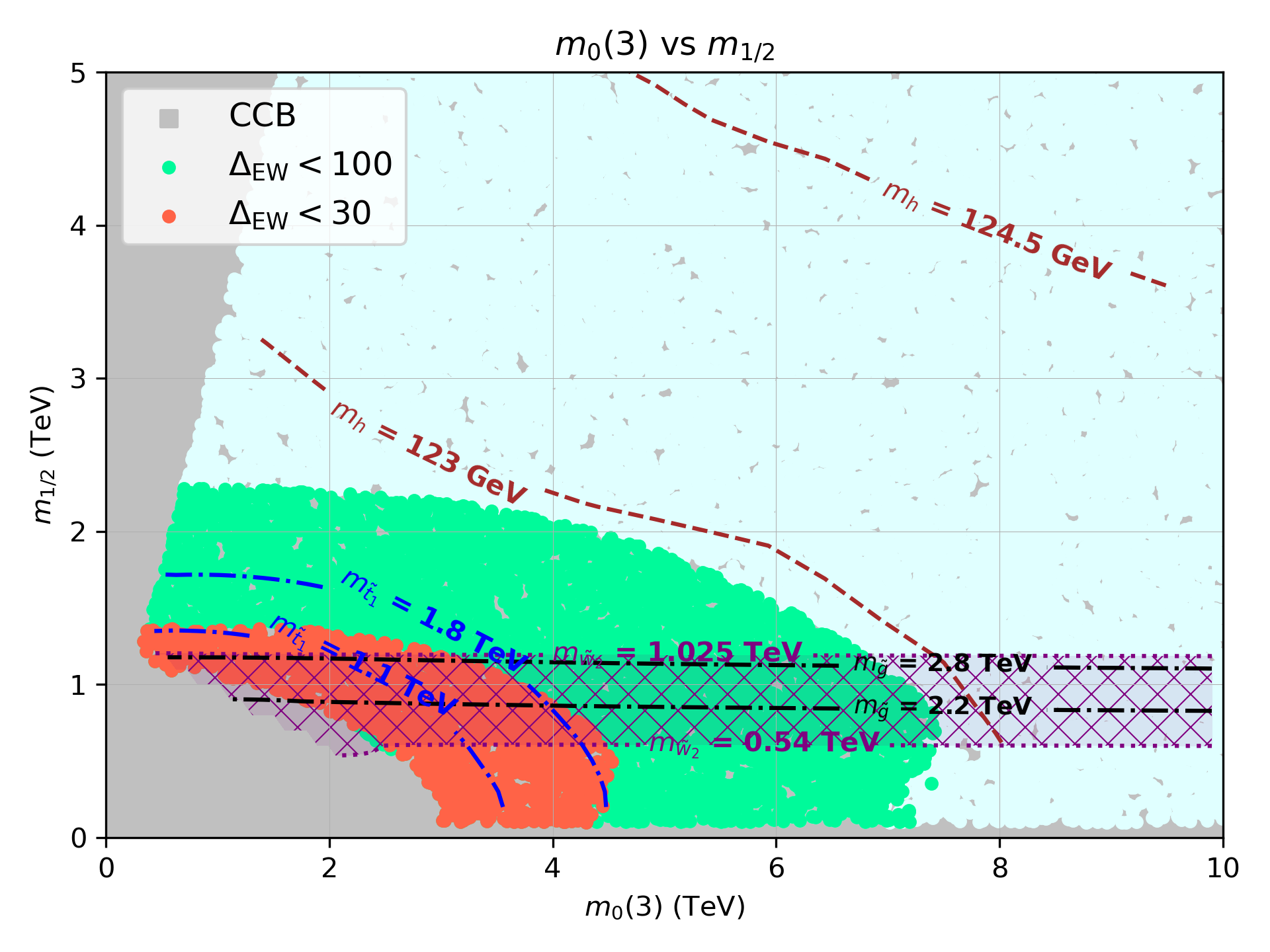}}\\
        {\includegraphics[height=0.28\textheight]{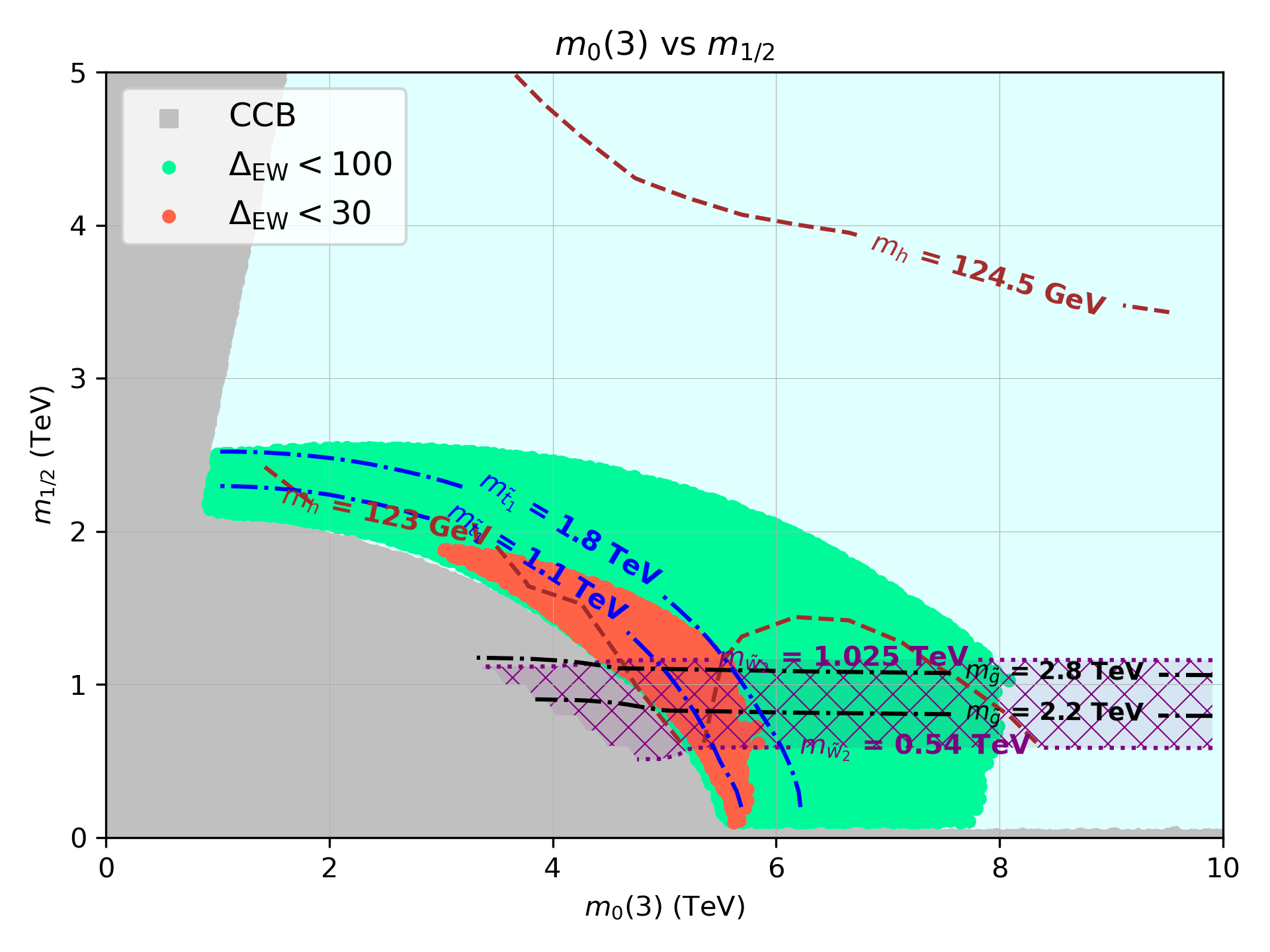}}\\
        {\includegraphics[height=0.28\textheight]{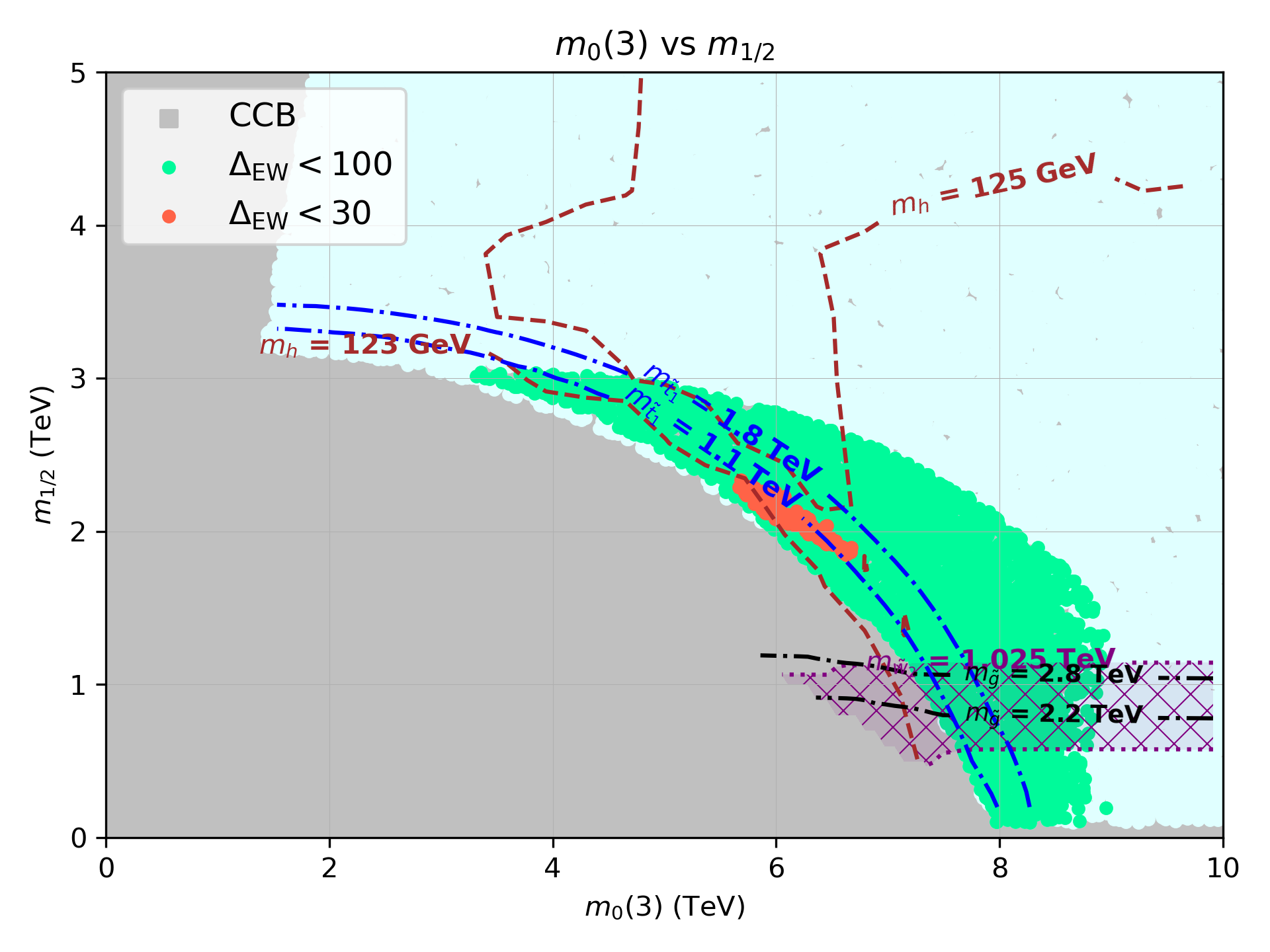}}
        \caption{Plot of NUHM3 model $m_0(3)$ vs. $m_{1/2}$
          plane for {\it a}) $m_0(1,2)=10$ TeV, {\it b}) $m_0(1,2)=20$ TeV and  {\it c}) $m_0(1,2)=30$ TeV. For each frame, we take
          $\mu =200$ GeV, $m_A=2$ TeV, $\tan\beta =10$ and $A_0=-m_0(3)$.
      \label{fig:nuhm3}}
\end{figure}

An advantage of complete models is that one can display the parameter space in terms of a few quantities and find correlations amongst various observables, be they flavor constraints, Higgs mass or collider bounds or projections for various sparticle and Higgs production cross sections.
With decoupled first/second generation scalars, then much of the action can be displayed in the $m_0(3)$ vs. $m_{1/2}$ parameter plane (analogous to the old CMSSM/mSUGRA plots in the $m_0$ vs. $m_{1/2}$ plane) for natural values of other parameters. Thus, in Fig. \ref{fig:nuhm3} we show this plane for
$\mu =200$ GeV, $m_A=2$ TeV, $\tan\beta =10$, $A_0=-m_0(3)$ and 
{\it a}) $m_0(1,2)=10$ TeV, {\it b}) $m_0(1,2)=20$ TeV and 
{\it c}) $m_0(1,2)=30$ TeV.

The first item of note is that the lower-left corner of each frame is
actually excluded by the presence of charge-or-color breaking (CCB) minima in the scalar potential (gray regions). This occurs when the stop soft term $m_{U_3}^2$ is driven negative by 2-loop RGE effects which become large under very heavy first/second generation sfermion masses\cite{Arkani-Hamed:1997opn,Baer:2010ny,Baer:2024hpl}. 
To offset this effect, one must go to large values of $m_0(3)$ or large $m_{1/2}$, where the heavier
gluino mass helps drive $m_{U_3}^2$ to large values. This helps explain
why SUSY wasn't discovered in the initial LHC runs since the lighter
gluino and third generation scalar mass regime has non-viable CCB minima.
This region grows in area as $m_0(1,2)$ increases, as shown by comparing the various frames in the figure.

We also see from the figure that the light Higgs mass $m_h$ is rather heavy due to large $A$-terms which are generated in SUGRA models. In fact, the bulk of the planes shown have the light Higgs mass values in accord with 
ATLAS/CMS measurements.

The natural regions of the parameter planes are listed by green (with 
$\Delta_{EW}<100$) and red (with $\Delta_{EW}<30$). 
The most natural regions lie adjacent to the CCB region, where top-squark soft masses are driven small-- but not negative-- resulting in small
values of $\Sigma_u^u(\tst_{1,2})$. 
In this region, we also show contours 
of light top-squark mass $m_{\tst_1}=1.1$ TeV (current limit from simplified model searches) and $1.8$ TeV (the HL-LHC $5\sigma$ expected reach for top-squarks as found in Ref. \cite{Baer:2023uwo}).\footnote{
For an earlier study, see Ref. \cite{Bai:2012gs}.
} 
What is notable about this result for HL-LHC (3000 fb$^{-1}$ at $\sqrt{s}=14$ TeV) is that the reach contours cover the entire natural parameter space! 
In the NUHM4 model, heavier top squarks with masses up to $\sim 2.5$ TeV can occur, but usually in models where $m_0(1,2)$ is much smaller and of order $\sim m_0(3)$ so that the SUGRA decoupling solution to the SUSY flavor/CP problems is not engaged. 
Thus, we expect that top-squark pair production with $m_{\tst_1}\sim 1-2$ TeV provides a lucrative avenue for SUSY discovery in the most plausible of SUSY models: gravity-mediation models which are natural, and which solve the SUSY flavor/CP problems via decoupling. 

\begin{figure}[htb!]
\centering
    {\includegraphics[height=0.5\textheight]{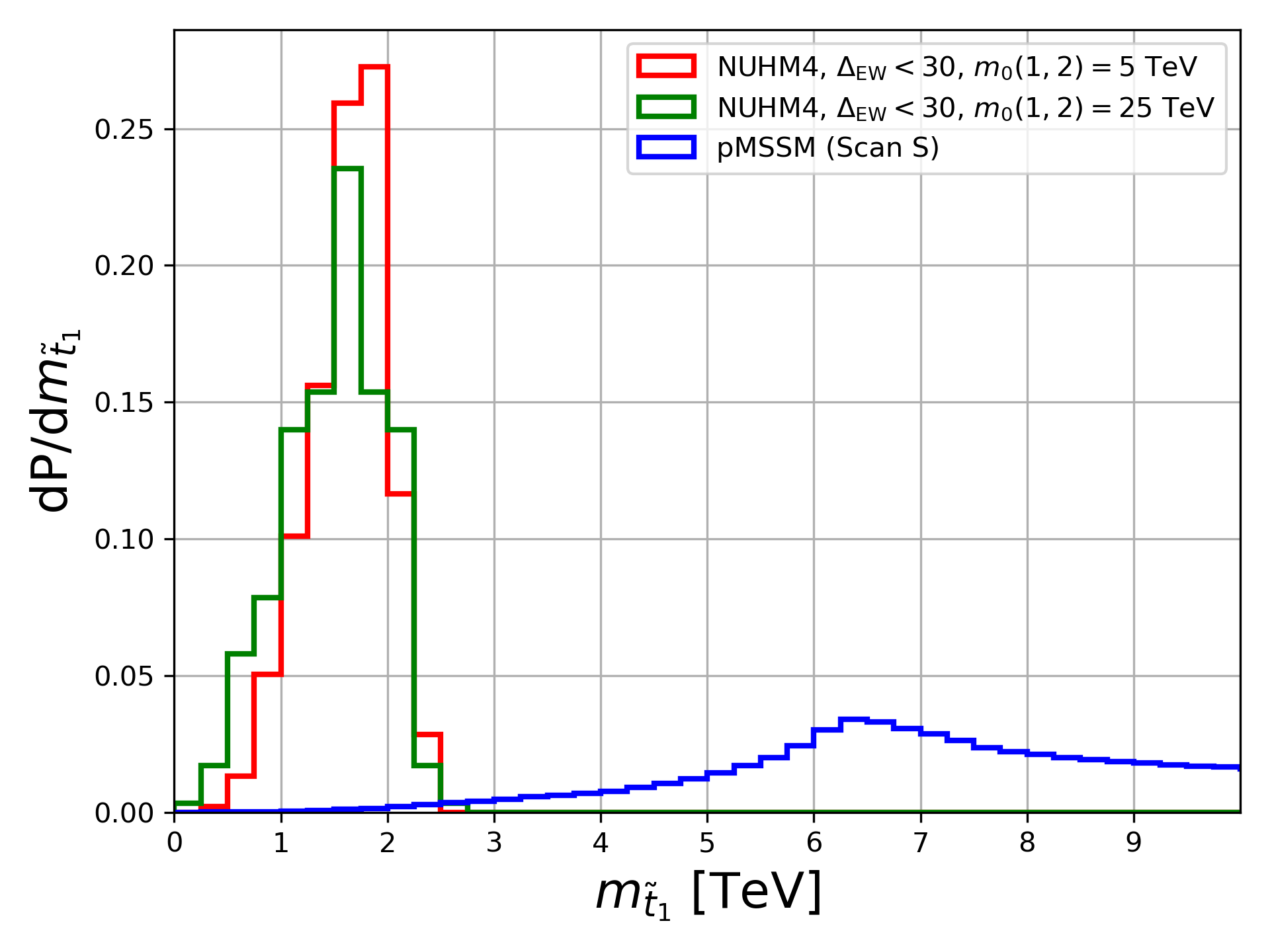}}
\caption{Probability distribution of expected light top-squark $m_{\tst_1}$
from the pMSSM scan {\bf S} (blue), from the string landscape but with 
$m_0(1,2)=5$ TeV (red) and landscape with $m_0(1,2)=25$ TeV (green).
      \label{fig:mt1}}
\end{figure}

The expected probability distributions for top-squark masses are shown in
Fig. \ref{fig:mt1}.  The blue histogram is the expected distribution from
pMSSM scan {\bf S} and extends over a hugs range (due to the large scan 
{\bf S} parameter range). This is to be compared to the string
landscape scan where $m_0(1,2)$ is fixed to 5 TeV (red) or 25 TeV (green).
The softening of the $m_0(1,2)=25$ TeV distribution compared to the
$m_0(1,2)=5$ TeV distribution is noticeable and is mainly a two-loop RG
effect which suppresses $m_0(3)$\cite{Baer:2024hpl} and pushes the stop mass
to more accessible values for HL-LHC searches.
The top-squark signal should complement the LHC soft-dilepton searches arising from higgsino pair production\cite{Baer:2011ec,Baer:2014kya,Baer:2021srt} 
which extend to $m(higgsino)\sim 250-300$ GeV\cite{Baer:2021srt}.

We also show in Fig. \ref{fig:nuhm3} contours of gluino mass $m_{\tg}=2.2$ TeV (current LHC simplified model limits) and $m_{\tg}=2.8$ TeV ($5\sigma$ 
reach limit of HL-LHC\cite{Baer:2016wkz}). The LHC gluino reach covers a limited portion-- but far from all-- of natural SUSY parameter space.

\section{Conclusions}
\label{sec:conclude}

In this paper we have critically examined the process of interpreting LHC SUSY search results within the framework of the pMSSM model, a 19 parameter weak scale SUSY model which discards RG running. 
The object of pMSSM interpretations is to invoke some sort of generality in SUSY search results. We argue that this is basically replacing theoretically well-motivated  constraints with a different set of unmotivated/unlikely prejudices. 
The scans over 19 weak scale parameters places almost all pMSSM models
into rather implausible territory, theoretically. These include 1. lack of RG evolution which is motivated by the successes of gauge coupling unification and radiative EWSB via the large top Yukawa coupling, 2. 
lack of matter unification even though fields of each generation fit snugly into the 16-dimensional spinor of $SO(10)$, 3. gaugino mass unification which is even more highly motivated nowadays (via the largish value of 
$m_h$ which requires large $A$-terms for natural SUSY models, which in turn implies singlet hidden sector SUSY breaking rather than charged SUSY breaking which provides for some sequestering), 4. naturalness via low $\Delta_{EW}$, which can be interpreted as to how well a given SUSY model predicts the weak scale at $m_{weak}\sim m_{W,Z,h}\sim 100$ GeV, 5. a decoupling solution to the SUSY flavor/CP problems which requires first/second generation matter scalars in the 10-40 TeV range (whereas pMSSM invokes a degeneracy solution in conflict with SUGRA) 6. artificial scan limits and scanning techniques which when changed alters pMSSM interpretations based on fraction of models surviving constraints and 
7. the often invoked constraints from WIMP/lightest neutralino production as all of the dark matter (almost always violated by stringy moduli, axion/axino/saxion presence in early universe etc.).

Instead, present LHC data are consistent with gravity-mediated SUSY
breaking in the form of the NUHM4 model with decoupled first/second generation scalars. In these models, natural portions of parameter 
space are easily found with $m_h\sim 125$ GeV. The decoupled first/second generation sfermions drive top-squark soft masses to live adjacent to CCB minima regions (which is sometimes called ``living dangerously''\cite{Arkani-Hamed:2005zuc}). 
In this parameter space, light top squarks seem within range of HL-LHC searches. Discovery is then expected in this channel at HL-LHC which can augment or even supersede discovery in higgsino or wino pair production searches\cite{Baer:2025zqt}. 
This is highly encouraging for Run 3 and HL-LHC searches
for SUSY, making for lucrative prospects for LHC SUSY discovery.

{\it Acknowledgements:} 

We thank X. Tata for collaboration during the early stages of this work.
VB gratefully acknowledges support from the William F. Vilas estate.
HB and KZ gratefully acknowledge support from the Avenir Foundation.


\bibliography{hgsno}
\bibliographystyle{elsarticle-num}

\end{document}